\RequirePackage{fix-cm}
\documentclass[prb,amsmath,amssymb,twocolumn,superscriptaddress]{revtex4-2}          
\usepackage{graphicx}
\usepackage[caption=false,justification=justified]{subfig}  
\usepackage{amssymb}
\usepackage{multirow}
\usepackage{enumitem}
\usepackage{siunitx}
\usepackage{booktabs}
\usepackage[dvipsnames,table,xcdraw]{xcolor}
\usepackage{etoolbox}

\newcommand{\vf}{\varphi}
\newcommand{\be}{\begin{equation}}
\newcommand{\ee}{\end{equation}}
\newcommand{\bea}{\begin{eqnarray}}
\newcommand{\eea}{\end{eqnarray}}
\renewcommand{\o}{\omega}

\newcommand{\nn}{\nonumber}
\newcommand{\ra}{\rangle}
\newcommand{\la}{\langle}
\newcommand{\Tr}{\textrm{Tr}}
\usepackage{amsmath}
\usepackage{hyperref}
\hypersetup{colorlinks=true,
	linktocpage=true,
	breaklinks=true,
	bookmarksnumbered,
	bookmarksopen=true,
	bookmarksopenlevel=1,
	hypertexnames=true,
	pdfhighlight=/0,
	urlcolor=black!38!red,
	linkcolor=MidnightBlue!60!Black,
	citecolor=PineGreen!60!Black
}
\usepackage{url}
\usepackage[utf8x]{inputenc}

\newcommand{\ped}[1]{_{\text{#1}}}
\newcommand{\api}[1]{^{\text{#1}}}
\newcommand{\lf}{\left}
\newcommand{\rg}{\right}
\DeclareUnicodeCharacter{8208}{ }
\begin{document}

\title{A hybrid ferromagnetic transmon qubit: circuit design, feasibility and detection protocols for magnetic fluctuations}
	\author{Halima Giovanna Ahmad}%
	\email{halimagiovanna.ahmad@unina.it}
	\affiliation{Dipartimento di Fisica "Ettore Pancini", Universit\`a di Napoli “Federico II”, Monte S. Angelo, I-80126 Napoli, Italy}%
	\affiliation{Seeqc, Strada Vicinale Cupa Cinthia, 21, I-80126 Napoli, Italy}%
	\affiliation{CNR-SPIN, UOS Napoli, Monte S. Angelo, via Cinthia, I-80126 Napoli, Italy}
	\author{Valentina Brosco}%
	\affiliation{Institute for Complex Systems, National Research Council and Dipartimento di Fisica, Universit\`a "La Sapienza", P.le A. Moro 2, 00185 Rome, Italy}
	\affiliation{Research Center Enrico Fermi, Via Panisperna 89a, 00184 Rome, Italy}%
	\author{Alessandro Miano}
	\altaffiliation[Current affiliation:]{ Department of Applied Physics, Yale University, New Haven, Connecticut 06520, USA}%
	\affiliation{Dipartimento di Fisica "Ettore Pancini", Universit\`a di Napoli “Federico II”, Monte S. Angelo, I-80126 Napoli, Italy}
	\author{Luigi Di Palma}%
	\affiliation{Dipartimento di Fisica "Ettore Pancini", Universit\`a di Napoli “Federico II”, Monte S. Angelo, I-80126 Napoli, Italy}
	\author{Marco Arzeo}%
	\affiliation{Seeqc, Strada Vicinale Cupa Cinthia, 21, I-80126 Napoli, Italy}%
	\author{Domenico Montemurro}%
	\affiliation{Dipartimento di Fisica "Ettore Pancini", Universit\`a di Napoli “Federico II”, Monte S. Angelo, I-80126 Napoli, Italy}%
	\author{Procolo Lucignano}%
	\affiliation{Dipartimento di Fisica "Ettore Pancini", Universit\`a di Napoli “Federico II”, Monte S. Angelo, I-80126 Napoli, Italy}%
	\author{Giovanni Piero Pepe}%
	\affiliation{Dipartimento di Fisica "Ettore Pancini", Universit\`a di Napoli “Federico II”, Monte S. Angelo, I-80126 Napoli, Italy}%
	\author{Francesco Tafuri}%
	\affiliation{Dipartimento di Fisica "Ettore Pancini", Universit\`a di Napoli “Federico II”, Monte S. Angelo, I-80126 Napoli, Italy}%
	\affiliation{CNR - Istituto Nazionale di Ottica (CNR-INO), Largo Enrico Fermi 6, 50125 Florence, Italy}
	\author{Rosario Fazio}%
	\affiliation{Abdus Salam ICTP, Strada Costiera 11, I-34151 Trieste, Italy }
	\affiliation{Dipartimento di Fisica "Ettore Pancini", Universit\`a di Napoli “Federico II”, Monte S. Angelo, I-80126 Napoli, Italy}%
	\author{Davide Massarotti}%
	\affiliation{Dipartimento di Ingegneria Elettrica e delle Tecnologie dell’Informazione$,$ Università degli Studi di Napoli Federico II$,$ via Claudio$,$ I-80125 Napoli$,$ Italy}
	\affiliation{CNR-SPIN, UOS Napoli, Monte S. Angelo, via Cinthia, I-80126 Napoli, Italy}

\begin{abstract}
We propose to exploit currently available tunnel ferromagnetic Josephson junctions to realize a hybrid superconducting qubit. We show that the characteristic hysteretic behavior of the ferromagnetic barrier provides an alternative and intrinsically digital tuning of the qubit frequency by means of magnetic field pulses. To illustrate functionalities and limitation of the device, we discuss the coupling to a read-out resonator and the effect of magnetic fluctuations. The possibility to use the qubit as a noise detector and its relevance to investigate the subtle interplay of magnetism and superconductivity is envisaged.
\end{abstract}


\maketitle

\section{Introduction}
\label{intro}

Superconducting qubits are among the most promising paradigms in quantum computation. A wealth of successful experiments proved how efficiently these devices can be manipulated and read-out by commercial electronics, how flexible is their Hamiltonian, how accurate is the control over their quantum state~\cite{Clarke2008,Devoret2013,Kockum2019,Krantz2019,DiCarlo2009}. More importantly, the steady progress of nanofabrication techniques and circuit's design brought a strong enhancement of qubit's coherence time that made possible performing practical quantum algorithms~\cite{Clarke2008,Devoret2013,Kockum2019,Krantz2019,DiCarlo2009}. The search for combinations of novel materials~\cite{Feofanov2010,Oliver2013,Lee2019,Place2021}, circuital designs~\cite{DiCarlo2009,Chen2014,Krantz2019,Kjaergaard2020} and new protocols for qubits encoding~\cite{McDermott2018,Mukhanov2019} is a very active field of research. The key role of the Josephson junctions in superconducting qubits has also promoted novel efforts for a better understanding their microwave properties, their electrodynamics parameters, and thus their non-linear behavior for the development of alternative approaches for the control and tunability of their functions.

Superconducting quantum circuits have almost exclusively relied on aluminum-aluminum oxide-aluminum (Al/AlO${\ped x}$/Al) tunnel Superconductor/Insulator/Superconductor (SIS) Josephson junctions (JJs)~\cite{Krantz2019,Kjaergaard2020}.  However, many exciting phenomena and functionalities can be accessed by exploiting unconventional superconducting systems. This has to be meant not only as a rush for the best qubit candidate, but also as an advance towards a better understanding and control of a Josephson-based quantum circuit. It is indeed of fundamental importance to explore novel hybrid quantum devices for enhancing both the capabilities of the superconducting electronics~\cite{Krantz2019,Kjaergaard2020} and the understanding of the exotic phenomenology that can arise in hybrid unconventional superconducting devices. As an example, in the specific case of tunable transmon qubits~\cite{Koch2007}, which tipically use external flux-fields to change the qubit frequency, hybrid superconductor-semiconductor structures~\cite{Barthel2009,Larsen2015,DeLange2015,Wiedenmann2016,Manousakis2017,Gul2018,Kroll2018,Casparis2018,Kunakova2020} have been used to enable voltage-tunable transmons (gatemons), in order to provide an alternative tuning of the qubit frequency without introducing flux-noise~\cite{DeLange2015}.

In this work, we focus on another special class of unconventional Josephson devices that use ferromagnetic barriers (SFS JJs). The competition between the superconducting and the ferromagnetic order parameters in these systems allows to build JJs with an intrinsic phase-shift of $\pi$~\cite{Buzdin2005,Minutillo2021}, providing $\pi$-components and \emph{quiet} qubits~\cite{Feofanov2010,Kawabata2006}. SFS JJs in superconducting circuits have been mostly used as passive elements~\cite{Feofanov2010} and they have not been considered up to now in the realization of quantum circuits, because of their intrinsic high quasi-particle dissipation~\cite{Buzdin2005,Eschrig2008,Cirillo2011,Blamire2014,Dahir2019}. This dissipation derives from the metallic nature of standard ferromagnetic  barriers, which unavoidably compromises the qubit state measurement and its performances~\cite{Serniak2018,Bilmes2020}. However, advances in coupling both ferromagnetic layers with insulating barriers inside the JJ (SIsFS or SIFS JJs)~\cite{Weides2006,Larkin2012,Vernik2013,Caruso2018,Parlato2020,Ryazanov2021} and the ability to exploit intrinsic insulating ferromagnetic materials (SI${}\ped f$S JJs)~\cite{Senapati2011,Pal2014,Massarotti2015,Caruso2019,Ahmad2020,Ahmad2022} allow to engineer ferromagnetic JJs characterized by high values of the quality factors and low quasiparticles dissipation~\cite{Ahmad2020}. Such tunnel-SFS JJs offer additional functionalities not only in superconducting classical circuits, but also in quantum architectures~\cite{Feofanov2010,Ahmad2020}. 

We here present a proof of concept study of an innovative protocol for qubit frequency tuning  relying on  state-of-the art tunnel-SFS JJs. Specifically, we focus on a transmon design featuring a  SIS JJ  and a SFS JJ inside a SQUID loop capacitively coupled to a superconducting read-out resonator~\cite{Koch2007}. We name this transmon employing ferromagnetic junctions \emph{ferro-transmon}. We discuss the possibility to provide: (i) a digital tuning of the qubit frequency exploiting the hysteretic nature of the F barrier, and (ii) a novel platform for the study of magnetization dynamics and fluctuations occurring in SFS JJs on a quantum-coherent scale. On one hand, the realization of digital read-out and tunability schemes for superconducting qubits may have a strong impact on the scalability of superconducting quantum systems~\cite{Mukhanov2019}. On the other hand, the possibility to probe magnetic noise fluctuations~\cite{Blamire2014,Cascales2019,Ahmad2020} provides an additional spectroscopic tool for the barrier dynamics and it may be of crucial relevance for the understanding of the unconventional superconducting transport mechanisms that occur at the S/F interface, including spin-triplet transport~\cite{Eschrig2008,Ahmad2022} and inverse proximity effect~\cite{Dahir2019,Satariano2021}.

The paper is organized as follows. In Sec.~\ref{highlights}, we discuss the working-principle behind the ferro-transmon, highlighting advantages, disadvantages and open issues. In Sec.~\ref{feasibility}, we discuss the feasibility of the device:  we estimate the ferro-transmon parameters, including the qubit frequency, the ferro-transmon tunability and the readout resonator dispersive shift as a function of the critical current of the SFS JJ. We compare the obtained results with those reported in literature for non-magnetic qubits~\cite{Krantz2019,Larsen2015}. In Sec.~\ref{dissipation}, we present a qualitative analysis of the dissipation mechanisms, focusing  in particular on the impact of SFS JJ's  magnetization fluctuations. Finally, we discuss a  protocol for the study of magnetization noise spectra relying on tunability of the couplings between the ferro-transmon and different noise sources.

\section{Ferro-transmon: a digitally tunable superconducting qubit}
\label{highlights}

\begin{figure*}[t!]
	\subfloat[][]{\includegraphics[scale=0.28]{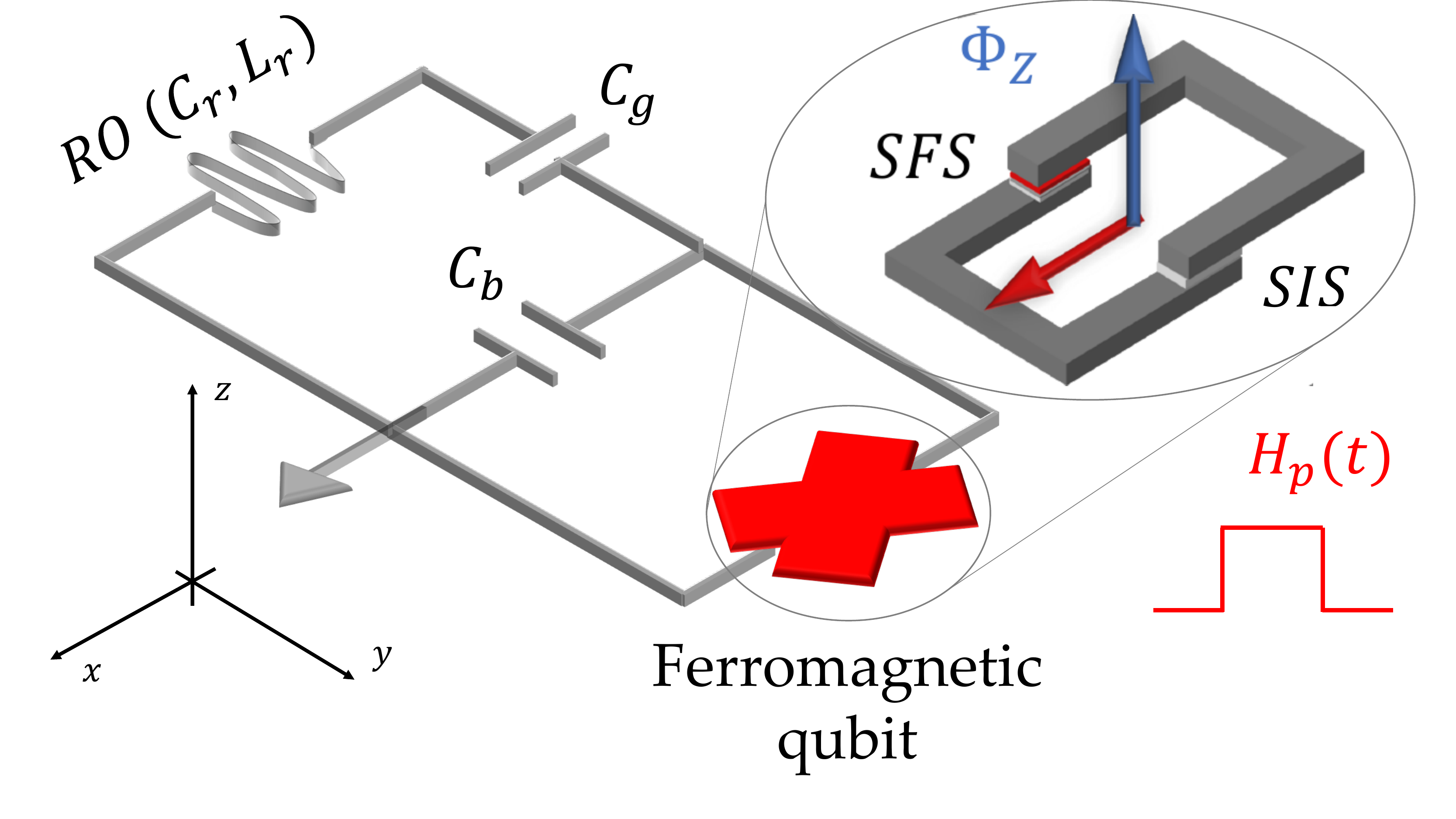}}
	\subfloat[][]{\includegraphics[scale=0.22]{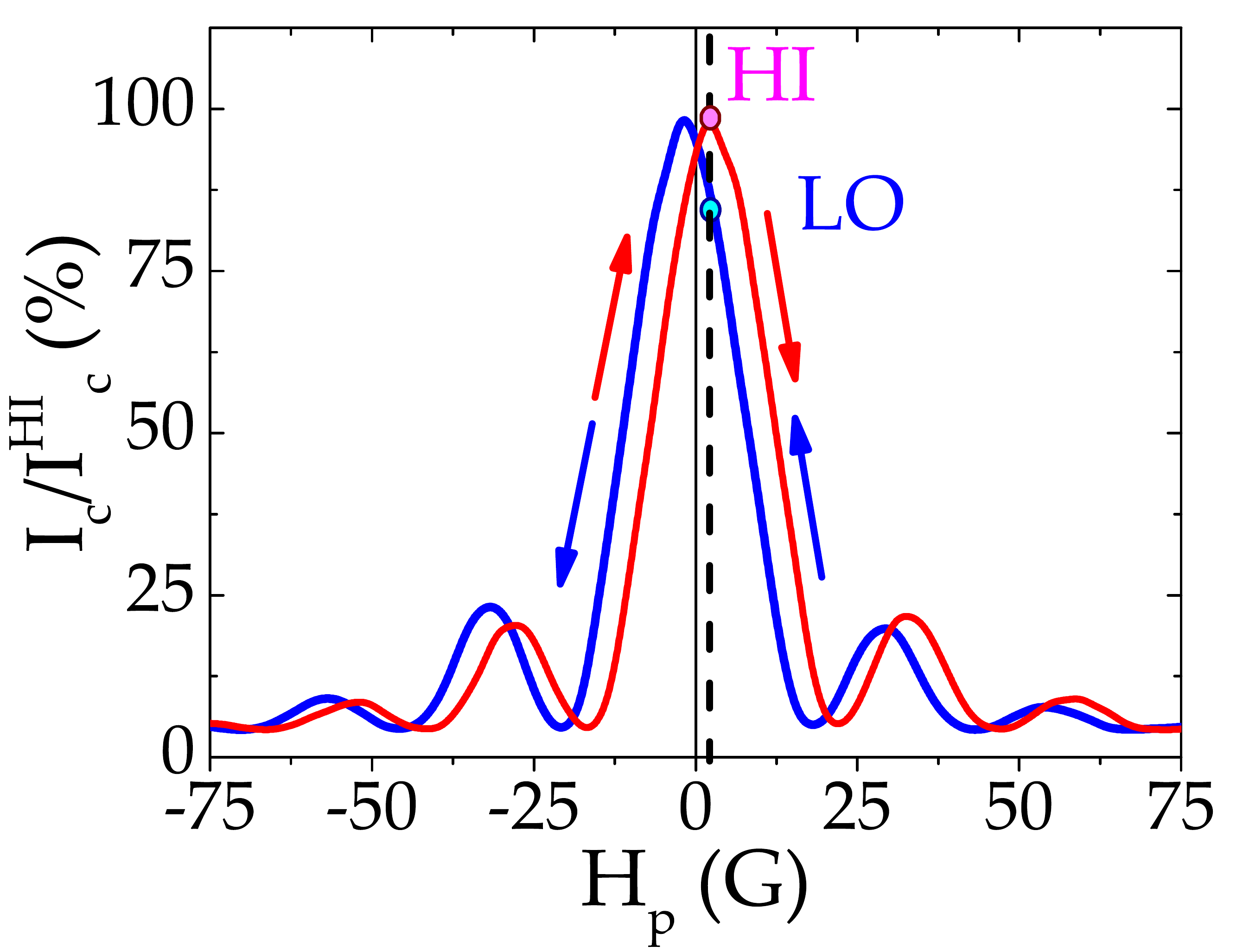}}\\
	\subfloat[][]{\includegraphics[scale=0.22]{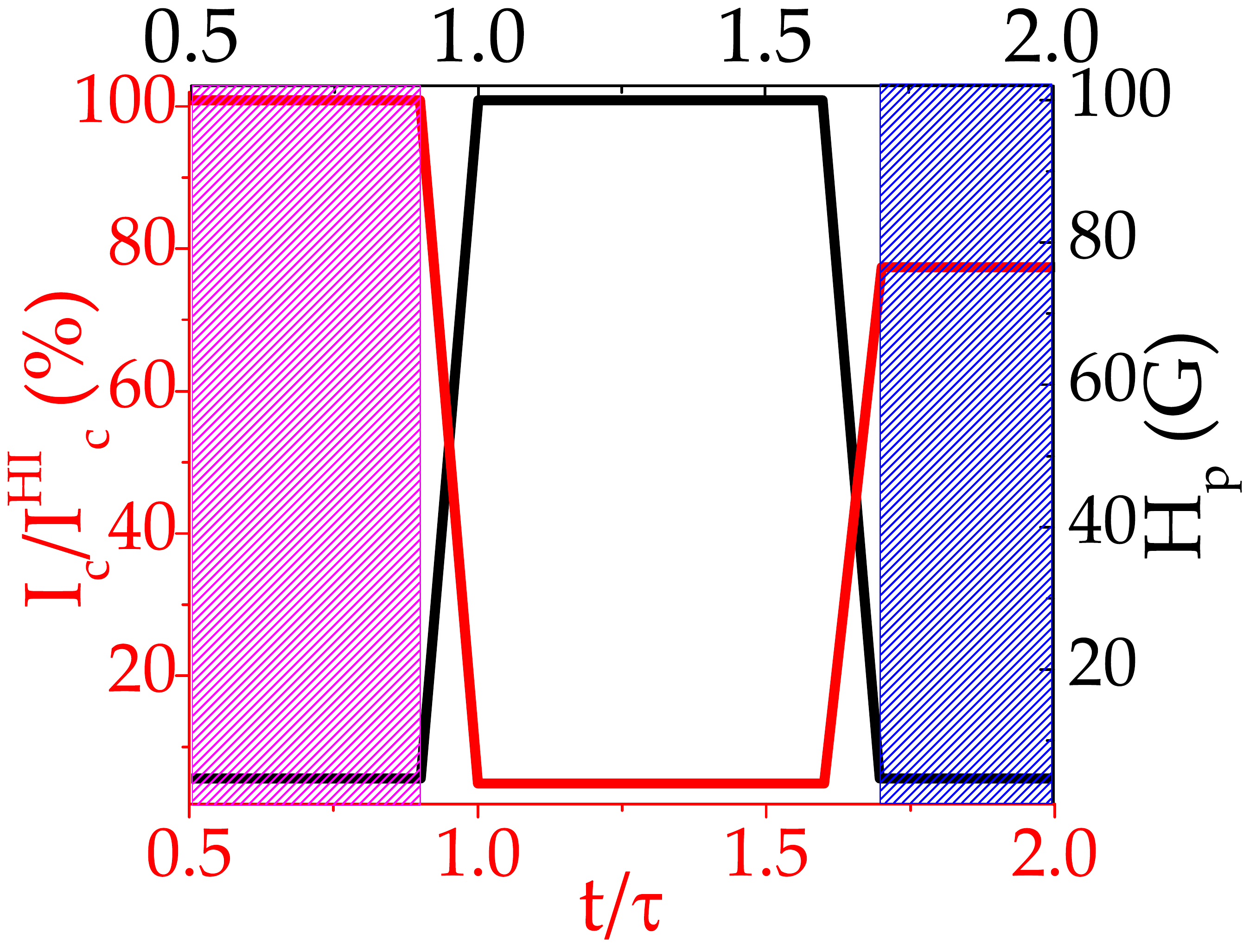}}\qquad\qquad\qquad\qquad\qquad\qquad
	\subfloat[][]{\includegraphics[scale=0.35]{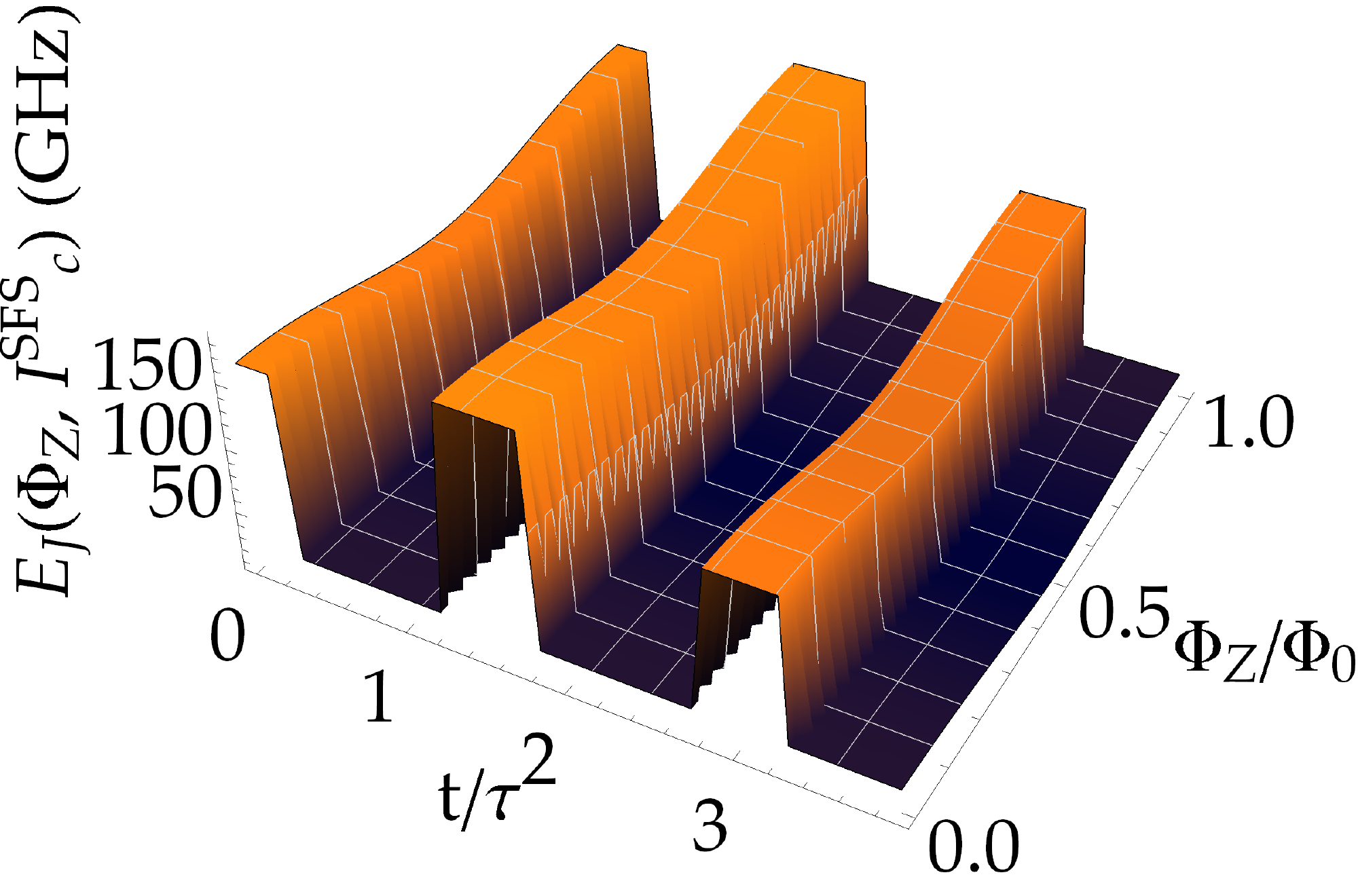}}\\
	\caption{In (a), ferro-transmon circuit design: the read-out (RO) resonator is capacitively coupled to the qubit through $C\ped g$. The qubit is schematized as a hybrid SQUID in parallel with a bias capacitor $C\ped b$. In the SQUID, there are a standard tunnel SIS JJ and a tunnel ferromagnetic SFS JJ. Blue and red arrows indicate magnetic fields applied along the $z$-axis and $x$-axis, respectively. In (b), $I\ped c(H)$ modulation in a tunnel SI${}\ped f$S JJ with a $\SI{3.5}{\nm}$ GdN barrier and area of $\SI{49}{\micro\m^2}$, normalized to the maximum $I\ped c\api{HI}=\SI{350}{\nA}$. Blue and red curves in (b) refer to the down and up magnetic field pattern, respectively.  We highlight in each plot the low- and high-$I\ped c$ level states (LO and HI) and the working point (dashed black line). In (c), example of a magnetic field pulse sequence $H\ped p$ (black line) and maximum digital tuning of the critical current $I\ped c/I\ped c\api{HI}$ (red line) for the JJ in (b). The blue and magenta dashed boxes refer to LO and HI levels, respectively. The time in (c) is normalized to a magnetic pulse timescale $\tau$ (see the text). In (d), calculated total Josephson energy $E\ped J$ of a hybrid SQUID composed of a SIS JJ with $E\ped J\api{SIS}=\SI{10}{\giga\hertz}$, {\sl i.e.} $I\ped c\api{SIS}\sim \SI{30}{\nano\ampere}$, and a SI${}\ped f$S JJ with GdN barrier. $I\ped c\api{SFS}$ is the critical current of the SFS JJ in the hybrid SQUID in (a), which is here fixed to $\SI{350}{\nano\ampere}$. The 3D-plot shows the dependence of $E\ped J$ on an external flux $\Phi\ped Z$ (in units of the quantum magnetic flux $\Phi_0$) and the magnetic field pulsed sequence $H\ped p(t)$ in (c), with time $t$ normalized to $\tau$. }
	\label{pattern}
\end{figure*}
We here discuss the main idea of the ferro-transmon circuit: digital tuning of the qubit frequency, based on the residual magnetization of a ferromagnet, can be achieved by integrating a tunnel-SFS JJ in the SQUID loop of a transmon circuit, schematically reported in Fig.~\ref{pattern} (a). The SQUID loop of the transmon is threaded as usual by an external flux $\Phi\ped Z$, which is directed along the $z$-axis. $\Phi\ped Z$ sets the standard cosinusoidal modulation of the SQUID Josephson energy~\cite{Barone1982,Tafuri2003}. In addition, in the ferro-transmon circuit a local magnetic flux $\Phi\ped L$ is applied along $x$ or $y$, {\sl i.e.} perpendicularly to the Josephson transport direction. 

The local field $\Phi\ped L$ is related to the magnetization loop of the ferromagnet $4\pi M(H\ped p)$ as~\cite{Larkin2012,Blamire2013,Vernik2013,Caruso2018,Parlato2020}
\begin{equation}\label{eq:phiL}
\Phi\ped L(H\ped p)=H\ped pd\ped ma+4\pi M(H\ped p)d\ped Fa,
\end{equation}
where $d\ped F$ is the thickness of the F barrier, $a$ is the lateral dimension of the tunnel-SFS JJ, $d\ped m$ is the effective magnetic length of the ferromagnetic barrier~\cite{Vernik2013} and $H\ped p$ is the field applied along $x$ or $y$~\cite{Larkin2012,Vernik2013}. While the pulsed $\Phi\ped L$ does not directly affect the behavior of the SIS junction, the critical current of the SFS JJ as a function of an external magnetic field $H\ped p$ along $x$ ($y$) follows a hysteretic Fraunhofer-like behavior, with a shift of the maximum of the $I\ped c(H)$ curve that depends on the residual magnetization of the F layer~\cite{Blamire2013}. As an example, in Fig.~\ref{pattern} (b) we report the $I\ped c(H)$ pattern measured at $\SI{10}{\milli\K}$ on a SI${}\ped f$S JJ with a $\SI{3.5}{\nm}$ GdN barrier and area of $\SI{49}{\micro\m^2}$, reported here as a reference. The S electrodes are made of NbN. Additional information about this device can be found in Refs.~\cite{Caruso2019,Ahmad2020,Ahmad2022}, while the fabrication procedure is reported in Refs.~\cite{Blamire2012,Pal2014,Massarotti2015}. A pulsed magnetic field as that represented in Fig.~\ref{pattern} (c) allows to switch the critical current of the SFS JJ between two discrete values, from now on defined as the low-level (LO) and the high-level (HI) states. As a consequence, also the Josephson energy can be digitally tuned. 
 
Formally, the dependence of the Josephson energy on $\Phi\ped Z$ and $\Phi\ped L$ reads as
\begin{align}
\label{EJ}
E\ped J(\Phi\ped{Z},\Phi\ped L)&={E\ped J}_{\Sigma}(\Phi\ped L)\cos\left(\pi \Phi\ped{Z}/\Phi_0\right)\cdot\\\notag
&\sqrt{1+d^2(\Phi\ped L)\tan^2\left( \pi\Phi\ped{Z}/\Phi_0\right)},
\end{align}
where we set $E_{ J\Sigma}(\Phi\ped L)={E\ped J}\api{SIS}+{E\ped J}\api{SFS}(\Phi\ped L)$ and we denote as $d(\Phi\ped L)$ the asimmetry parameter,
\be
\label{d}
d(\Phi\ped L)=\frac{E_{\rm J}\api{SIS}-E_{\rm J}\api{SFS}(\Phi\ped L)}{E_{\rm J}\api{SIS}+E_{\rm J}\api{SFS}(\Phi\ped L)}.
\ee
In Fig.~\ref{pattern} (d), we report the Josephson energy $E\ped J$ as a function of the external flux $\Phi\ped Z$ applied along the axis $z$ of the SQUID and as a function of the same pulsed magnetic field sequence in Fig.~\ref{pattern} (c) applied along $x$. In the ferro-transmon SQUID, we consider a SI${}\ped f$S JJ with GdN barrier as that in Fig.~\ref{pattern} (b), and a Josephson energy for the non-magnetic JJ of standard values commonly found in Al/Nb based transmon qubits, {\sl i.e.} of the order of ${E\ped J}\api{SIS}=\SI{10}{\giga\hertz}$~\cite{Walraff2004,Majer2007}. As one can observe, the application of the magnetic field along $x$ allows to tune $E\ped J$ between two discrete values: before the pulse, $I\ped c$ is in the HI state. At the end of the pulse $I\ped c$ reaches the LO state. By applying $\Phi\ped Z$ along the $z$-axis, instead, we observe the typical cosinusoidal modulation of $E\ped J$ given by the flux-tunability of the SQUID. In the specific case analyzed, $E\ped J(\Phi\ped{Z},\Phi\ped L)$ at $\Phi_0/2$ does not vanish. This tipically occurs when the asymmetry of the SQUID reaches $d\sim 1$. As a matter of fact, in the mentioned reference example reported in Fig.~\ref{pattern}, the JJs in the SQUID have critical currents that differ by a factor $10$, since ${I\ped c}\api{SFS}\sim\SI{350}{\nano\ampere}$ at dilution temperatures~\cite{Ahmad2022}, while ${I\ped c}\api{SIS}\sim \SI{30}{\nano\ampere}$. This also implies that far from the sweet-spots (multiple semi-integer of $\Phi_0$), the flux-noise sensitivity of the transmon qubit is strongly reduced~\cite{Koch2007,Hutchings2017}.

The possibility to employ a large variety of ferromagnetic materials gives the opportunity to identify and engineer magnetic field pulses without any limitation. For instance, in the framework of Nb or NbN technology, GdN-based JJs have shown critical current variations $\Delta I=(I\ped c\api{HI}-I\ped c\api{LO})/I\ped c\api{HI}$ of the order of $25\%$ (Fig.~\ref{pattern} (b)). Other ferromagnetic materials employed in Nb-based JJs, such as permalloy-based JJs~\cite{Parlato2020,Satariano2021} or palladium-iron PdFe-based JJs~\cite{Larkin2012,Vernik2013,Caruso2018} can give $\Delta I$ of the order of $40\%$. Particularly relevant for PdFe barriers in SIsFS JJs, $\Delta I$ can be also enhanced with the application of RF-fields~\cite{Caruso2018}, compatibly with standard microwave equipment. Moreover, $\Delta I$ can be adjusted by designing a pulsed field sequence that exploits minor magnetization loops so to reduce the hysteresis in the $I\ped c(H)$ modulation, and the separation between LO an HI critical current levels~\cite{Caruso2018,Parlato2020}. The magnitude of $H\ped p$ does not require to reach the saturation field of the ferromagnet, which may affect the performances of the device. The working point in Fig.~\ref{pattern} is chosen such to have the largest separation between the LO and the HI state, $\Delta I$. It is possible to engineer SFS JJs with finite $\Delta I$ at a working point corresponding to $H\ped p=0$~\cite{Ryazanov2021} thus avoiding the application of external magnetic fields that may be detrimental for qubit coherence, also by employing asymmetric minor loops.

The magnetic field pulses time in Fig.~\ref{pattern} is normalized to a general timing of the pulsed field sequence $\tau$, which strongly depends on both the magnetization dynamics time-scale $\tau\ped c$ and the switching-speed of the tunnel-SFS JJ into play. Thus, a careful choice of  the electrodynamics and magnetic properties of the tunnel-SFS JJ is fundametal to guarantee the largest speed of the digital tunability protocol. As an example, tipical magnetization dynamics occurs on a time-scale $\tau\ped c<\si{\nano\s}$~\cite{Dietricht2000}, which is far lower than the coherence times in transmon qubits range (some microsecond to hundreds of microseconds) and current state-of-the-art single- and two-qubit gate operations (tens to hundreds of nanoseconds)~\cite{DiCarlo2009,Chen2014,Krantz2019,Sung2021coupler}. The switching speed is defined by the $I\ped cR\ped N$ product of the tunnel-SFS JJ, which has already been demonstrated to be compatible with high-speed and energy-efficient superconducting digital circuits~\cite{Larkin2012,Vernik2013,Caruso2018,Parlato2020,Satariano2021}, such as Single-Flux Quantum logic electronics~\cite{Mukhanov2011,Mukhanov2019}. Specifically considering the NbN-GdN-NbN JJ in Fig.~\ref{pattern} (b), the switching speed at $\SI{10}{\milli\K}$ is about $\SI{35}{GHz}$, which corresponds to $\SI{28}{\pico\second}$. Particularly relevant for the purpose of integrating ferromagnetic materials in standard fabrication of transmon qubits without affecting their quality, in the case of SIsFS JJs the Josephson switching speed is only related to the SIs trilayer, provided that the intermediate s thickness $d\ped s$ exceeds the coherence lenght of the superconductor $\xi\ped s$~\cite{Bakurskiy2013}. In this case, the SIsFS JJ works as a series of a standard tunnel SIs JJ and the sFS JJ~\cite{Parlato2020,Satariano2021}. This ensures that: (i) the SIs JJ can be easily integrated in the transmon through standard fabrication procedures, while the F layer can be deposited afterwords without affecting the quality of the SIs trilayer, and (ii) the switching speed and the quality of the JJ are only related to the tunnel SIs JJ. The S electrode in the sFS JJ can be designed to work also as a current-line for the application of in-plane magnetic fields, which meets the request of a scalable device~\cite{Ryazanov2021}. 

While on one hand the simulatenous presence of an external flux field $\Phi\ped Z$ and a local pulsed flux $\Phi\ped L$ makes this system a useful playground with multiple knobs able to tune $E\ped J$, on the other hand the proposed design allows to tune $E\ped J$ only through $\Phi\ped L$ pulses. This means that instead of a hybrid SQUID in the qubit, it may be worth to explore also other circuital design in which a single tunnel-SFS JJ is used, thus completely removing the effect of additional flux-noise fluctuations. 

In the next section, we will discuss the role played by the electrodynamics parameters of the tunnel-SFS JJ and the capacitive couplings in the circuit design in the value of the qubit frequency, its tunability and the readout resonator dispersive shift. We focus on the estimation of the critical currents for the SFS JJs to build a reliable and measurable ferro-transmon device.
\begin{figure*}
	\subfloat[][]{\includegraphics[width=0.5\textwidth]{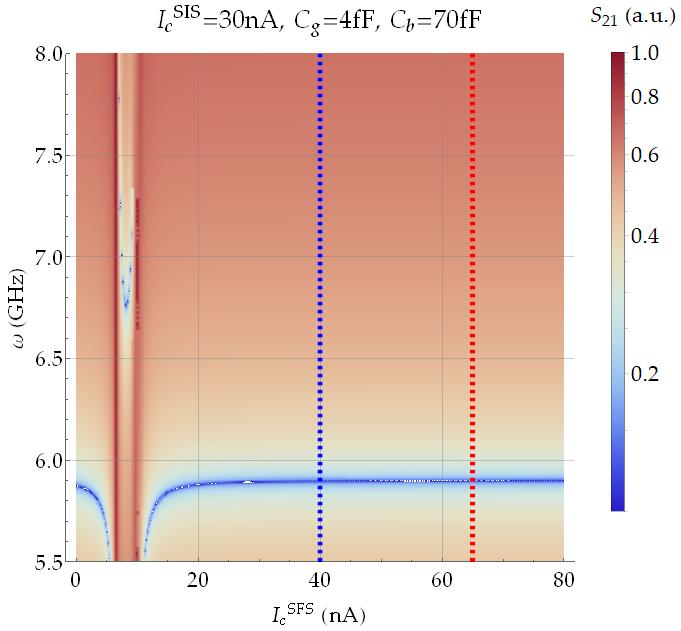}}
	\subfloat[][]{\includegraphics[width=0.5\textwidth]{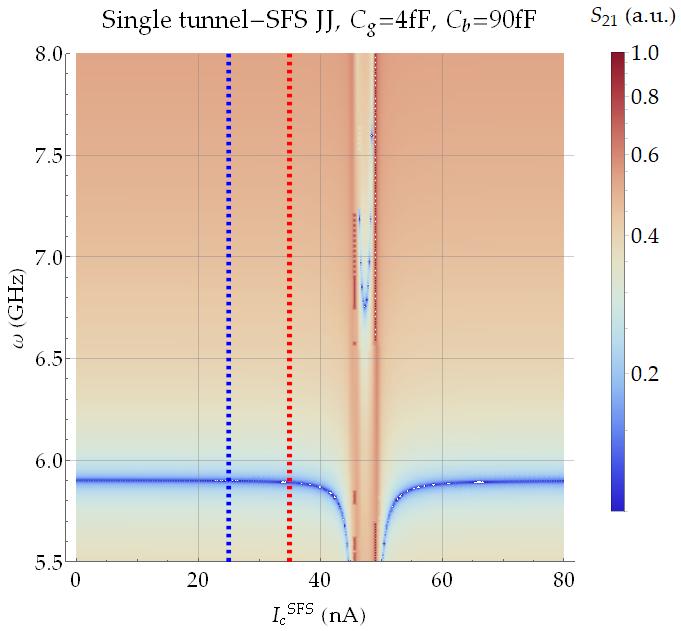}}\\
	\subfloat[][]{\includegraphics[width=0.5\textwidth]{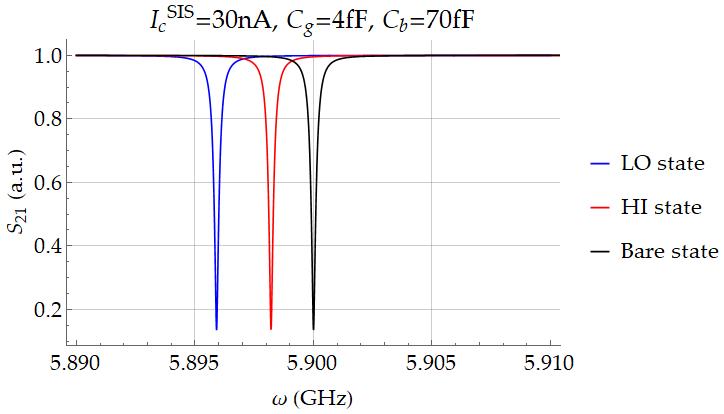}}
	\subfloat[][]{\includegraphics[width=0.5\textwidth]{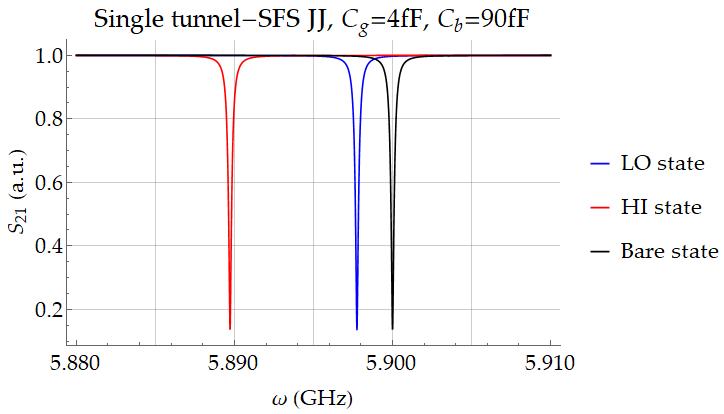}}\\
	\caption{Transmission parameter $S_{21}(\omega)$ in arbitrary units for the ferro-transmon read-out resonator as a function of $I\ped c\api{SFS}$, the critical current in the tunnel-SFS JJ. In (a) and (b), we explore the hybrid DC-SQUID configuration, with SIS JJ critical current $I\ped c\api{SIS}=\SI{30}{\nano\ampere}$, and the ferro-transmon configuration with a single tunnel-SFS-JJ, respectively. We fix $C\ped b=\SI{70}{\femto\farad}$ in (a) and $C\ped b=\SI{90}{\femto\farad}$ in (b), and $C\ped g=\SI{4}{\femto\farad}$ in both the configurations. The color-scale in (a) and (b) refers to the $S_{21}$ of a notch-type resonator in a log-scale~\cite{Probst2015}, with load and coupling quality factor $Q\ped l=1.7\cdot10^4$ and $Q\ped c=2\cdot10^4$, respectively, and no impedance mismatch at input and output port of the feedline. In (c) and (d), we show the readout resonator frequency dispersive shift in the high (low) level states (LO in blue and HI in red, respectively) of the SFS JJ for the configurations in panels (a) and (b), respectively. HI and LO levels in (c) and (d) are highlighted in panels (a) and (b) with the red and blue vertical dashed lines, respectively, and reported in Tab.~\ref{Tab:1} for completeness.}
	\label{final}
\end{figure*}

\section{Feasibility}
\label{feasibility}

The tunnel-SFS critical current $I\ped c\api{SFS}$ and its tuning with a local pulsed magnetic field $\Phi\ped L$ set the values of the ratio $E\ped J/E\ped c$ and the qubit frequency $\omega_{01}$, given by
\begin{equation}
\label{freq}
\omega_{01}(\Phi\ped{Z},\Phi\ped L)=\sqrt{8{E\ped c}_{\Sigma}E\ped J(\Phi\ped{Z},\Phi\ped L)}-{E\ped c}_{\Sigma},
\end{equation} 
as discussed in Ref.~\cite{Koch2007} and App.~\ref{App1}. As expected, when the ferro-transmon is coupled to a superconducting resonator, the pulsed magnetic field tuning significantly affects the effective qubit-resonator coupling and the electromagnetic response of the system. In this section, to assess the feasibility of the ferro-transmon and define the details on its circuital design, we study its elecromagnetic response in the two HI and LO states of the tunnel-SFS JJ, yielding $I\ped{c}\api{SFS}(\text{HI})$ and $I\ped{c}\api{SFS}(\text{LO})$.

The superconducting read-out resonator is designed as a $\lambda/4$-coplanar waveguide with bare-frequency $\omega\ped{RO}\api{bare}=\SI{5.9}{\giga\hertz}$ in a notch-type geometry~\cite{Probst2015}. The effective capacitance of the resonator in the lumped element approximation is $C\ped r=\SI{825}{\fF}$~\cite{Pozar2011}. This choice has been made in order to discuss the transmon read-out for standard parameters in literature~\cite{Blais2004,Krantz2019,Rigetti2012,Hutchings2017}. For the same reason, we keep ${E\ped c}_{\Sigma}\sim\SI{200}{\mega\hertz}$~\cite{Blais2004,Krantz2019,Rigetti2012,Hutchings2017}, where ${E\ped c}_{\Sigma}=e^2/(2C_{\Sigma})$ and $C_{\Sigma}$ is the total effective capacitance in Eq.~(\ref{total_capacitance}), which depends on the coupling capacitance with the read-out resonator $C\ped g$ and the qubit bias capacitor $C\ped b$ in Fig.~\ref{pattern} (a)~\cite{Koch2007}. Finally, the qubit-readout coupling $g$ for the first two levels of the transmon reads as~\cite{Koch2007},
\begin{equation}
g=g_{01}=\frac{e}{\hbar}\frac{C\ped g}{C_{\Sigma}}\sqrt{\frac{\hbar\omega\ped{RO}\api{bare}}{C\ped r}}\left(\frac{E\ped J}{8E\ped c}\right)^{1/4},
\end{equation} 
{\sl i.e.} it depends on the bare-resonator frequency $\omega\ped{RO}\api{bare}$ and the ratio between the coupling capacitance $C\ped g$ and the total transmon capacitance $C_{\Sigma}$~\cite{Koch2007}.

To explore the full range of available parameters, beside the case of a hybrid DC-SQUID dispersively coupled to a superconducting resonator, we here consider also a ferro-transmon made of a single tunnel-SFS JJ. In the latter case the tuning of the qubit frequency is achieved only through the magnetic pulsed field tuning.

A first estimation of the charging energy ${E\ped{c}}_{\Sigma}$, the ratio $E\ped J/E\ped c$, the qubit frequency $\omega_{01}$ and the read-out coupling $g$, as a function of $I\ped c\api{SFS}$ and standard capacitances in the circuit, suggests the following ranges of values for the tunnel-SFS JJ critical current: (i) for $I\ped c\api{SFS}$ ranging from $40$ to $\SI{65}{\nA}$ for the hybrid DC-SQUID configuration, once set ${E\ped c}_{\Sigma}=\SI{260}{\mega\hertz}$, i.\,e. $C\ped g=\SI{4}{\fF}$ and $C\ped b=\SI{70}{\fF}$, and $I\ped c\api{SIS}$ from $\SI{10}{\nA}$ to $\SI{30}{\nA}$; (ii) for $I\ped c\api{SFS}$ ranging from $25$ to $\SI{80}{\nA}$ for the single tunnel-SFS JJ, once set ${E\ped c}_{\Sigma}=\SI{200}{\mega\hertz}$, i.\,e. $C\ped g=\SI{4}{\fF}$ and $C\ped b=\SI{90}{\fF}$. These values guarantee a robust transmon regime, {\sl i.e.} with $E\ped J/E\ped c$ of the order of $50-100$~\cite{Koch2007}. Furthermore, as occurs in conventional transmon circuits based both on the Al or Nb technology, $\omega_{01}<\SI{10}{\giga\hertz}$, {\sl i.e.} easily detected using standard qubit measurement equipment~\cite{Nakamura2011,Krantz2019,Kim2021}. Charging energies above $\sim\SI{200}{\mega\hertz}$ ensure sufficently large anharmonicity to isolate a quantum two-level system~\cite{Nakamura2011,Krantz2019,Kim2021}.    

Compared to what shown in Fig.~\ref{pattern} (d), in which the amplitude of $E\ped J$ is of the order of some hundreds of gigahertz, {\sl i.e.} one order of magnitude larger than typical values in standard transmon devices, the reduction of $I\ped c\api{SFS}$ is a fundamental step in order to provide a reliable and measurable ferro-transmon device. While suitable parameters may in principle be obtained by using larger $I\ped c\api{SFS}$, the ${E\ped c}_{\Sigma}$ must be decreased far below $\SI{100}{\mega\hertz}$, thus unavoidably affecting the anharmonicity of the two-level system. However, compared to the drawbacks due to a reduction of ${E\ped c}_{\Sigma}$, the $I\ped c\api{SFS}$ can be reduced by keeping the critical current density $J\ped c\api{SFS}$ constant, and therefore reducing the area of the device, or by reducing $J\ped c\api{SFS}$ itself. 

For GdN-based JJs, for example, the $J\ped c\api{SFS}$ can be decreased by a factor $10$ by increasing the thickness of the GdN interlayer from $\SI{3.5}{\nm}$ to $\SI{4.0}{\nm}$, where typical values of $\SI{10}{\nano\ampere}$ have already been reported~\cite{Caruso2019,Ahmad2020}. For tunnel SIsFS JJs, $J\ped c\api{SFS}$ can be regulated changing the area of the tunnel SIs trilayer in the JJ~\cite{Parlato2020}. By using strong ferromagnets such as the permalloy, for example, the area of the JJ can be successfully scaled to dimensions of the order of some $\si{\micro \m^2}$~\cite{Parlato2020}, and in principle could be scaled in the sub-micrometer regime. The scaling of the area can not be accomplished, instead, on most of the soft ferromagnets used in SIsFS JJs, since they are often percolative systems or strongly dependent on a multi-domain configuration~\cite{Cullity2011,Bolginov2017}. However, in this case $J\ped c\api{SFS}$ may be reduced increasing the thickness of the insulating layer. 

Finally, let us refer to the analytical approach proposed by Koch et al. in Ref.~\cite{Koch2007} to give the read-out dispersive frequency shift $\chi$ as a function of the HI and LO level states of the tunnel-SFS JJ, for both the hybrid DC-SQUID and the single tunnel-SFS JJ ferro-transmon designs. In order to work in the dispersive regime, the coupling between the read-out resonator and the qubit $g$ in the transmon must satisfy the condition $g\ll\Delta$, where $\Delta=\omega\ped{RO}\api{bare}-\omega_{01}$ is the detuning between the read-out resonator frequency and the first-order transition frequency for the qubit~\cite{Koch2007}. Typical coupling factors range from $g\sim\SI{10}{\mega\hertz}$ to $\SI{100}{\mega\hertz}$~\cite{Koch2007,Krantz2016,Krantz2019}. In Fig.~\ref{final}, we report the ideal resonator transmission parameter $S_{21}(\omega)$~\cite{Probst2015}, as a function of $I\ped c\api{SFS}$,
\begin{equation}
S_{21}(\omega,I\ped c\api{SFS})=\left|1-\frac{Q\ped l/Q\ped ce^{i\eta}}{1+2iQ\ped l\left(\frac{\omega}{\omega\ped{RO}(I\ped c\api{SFS})}-1\right)}\right|,
\end{equation} 
where $\omega\ped{RO}(I\ped c\api{SFS})=\omega\ped{RO}\api{bare}+\chi(I\ped c\api{SFS})$, $Q\ped l$ is the resonator total quality factor and $Q\ped c$ is the coupling quality factor, fixed to $1.7\cdot10^4$ and $2\cdot 10^4$, respectively~\cite{Khanna1983,Palacios2008,Partanen2018}. $\eta$ is the impedance mismatch at the input and output ports of the coupled feedline~\cite{Probst2015}, which is here fixed to $0$. Also quality factors have been chosen to comply with standard parameters in literature~\cite{Khanna1983,Palacios2008,Probst2015,Partanen2018}.

We report in Fig.~\ref{final} (a) simulations for the hybrid DC-SQUID configuration with $I\ped c\api{SIS}=\SI{30}{\nA}$, while in (b) we focus on the single tunnel-SFS JJ configuration. There are specific $I\ped c\api{SFS}$ regions for which the coupling overcomes the dispersive regime (straddling regime~\cite{Krantz2019}), {\sl i.e.} around $\SI{10}{\nA}$ in panel (a) and around $\SI{50}{\nA}$ in panel (b). This limits the ferro-transmon operation. In order to fall in the transmon and in the dispersive regime with feasible qubit frequency values, and to be far from the straddling regime, for the hybrid DC-SQUID configuration the best HI and LO achievable values have to be $\SI{65}{\nA}$ and $\SI{40}{\nA}$, respectively. Vertical line-cuts related to these fixed $I\ped c\api{SFS}$ values are reported in panel (c), and compared with the transmission of the bare resonator. The same arguments can be given for the single tunnel-SFS ferro-transmon configuration in panel (b), in which we find $I\ped c\api{SFS}(\text{HI})=\SI{35}{\nA}$ and $I\ped c\api{SFS}(\text{LO})=\SI{25}{\nA}$. Vertical line-cuts related to these values are reported in panel (d). A summary of the ferro-transmon electrodynamics parameters for the two configurations shown in panels (c) and (d) are finally reported in Tab.~\ref{Tab:1}, where we collect $E\ped J/E\ped c$ ratio, $\omega_{01}$, its tunability $\Delta\omega_{01}=(\omega_{01}(\text{HI})-\omega_{01}(\text{LO}))/\omega_{01}(\text{HI})$ and the qubit-readout coupling $g$ for the HI and LO state.
	
All the calculated $I\ped c\api{SFS}$ values are compatible with current technologies, thus making feasible the ferro-transmon. As a matter of fact, in both the proposals, the ratio $E\ped J/E\ped c$ is comparable or larger than in typical transmon circuits. Also the qubit frequencies fall in the operational qubit frequency range~\cite{Hutchings2017}. Moreover, within the chosen circuital parameters, the resonator shifts $\chi$ is such to discriminate between the HI and LO state through the transmon read-out resonator in both the configurations. Most importantly, the qubit frequency tunability through a pulsed local magnetic field ranges from $\Delta\omega\ped Q\sim\SI{0.8}{\giga\hertz}$ for the single tunnel-SFS JJ to $\SI{1}{\giga\hertz}$ for the hybrid DC-SQUID configuration, as in typical flux-tunable transmons~\cite{Hutchings2017}. Such tunability range corresponds to $\Delta I\sim30\%$, which can be properly engineered even by exploiting minor magnetization loops~\cite{Ryazanov2012,caruso2018a,Parlato2020}. We stress that these parameters  can be further adjusted also by changing the resonator parameters and the capacitive elements in the circuit.  
\begin{table*}
	\caption{Summary of the ferro-transmon parameters analyzed in this work: the charging energy ${E\ped c}_{\Sigma}$, the percentual difference between the qubit frequencies in the HI (high) and LO (low) state $\Delta \omega_{01}$, the ratio $E\ped J/E\ped c$, the qubit frequency $\omega_{01}$, the dispersive shift $\chi$ and the read-out coupling $g$, obtained following the analytical approach proposed by Koch et al.~\cite{Koch2007}. All the parameters are calculated for the hybrid DC-SQUID ferro-transmon configuration with $I\ped c\api{SIS}=\SI{30}{\nano\ampere}$ (a), and for a ferro-transmon with a single SFS-JJ (b). Each of the configurations is characterized by $C\ped g=\SI{4}{\fF}$, while $C\ped b$ is $\SI{70}{\fF}$ for the former, and $\SI{90}{\fF}$ for the latter.}
	\label{Tab:1}
	\centering
	\begin{tabular}{lllllll}
		\hline\noalign{\smallskip}
		 & (a) & $I\ped c\api{SIS}=\SI{30}{\nA}$ & & (b) & Single tunnel-SFS JJ & \\
		\noalign{\smallskip}\hline\noalign{\smallskip}
		 & & HI : $I\ped c\api{SFS}=\SI{65}{\nA}$ & LO: $I\ped c\api{SFS}=\SI{40}{\nA}$ & & HI : $I\ped c\api{SFS}=\SI{35}{\nA}$ & LO: $I\ped c\api{SFS}=\SI{25}{\nA}$  \\
		\noalign{\smallskip}\hline\noalign{\smallskip}
		${E\ped c}_{\Sigma}$ $(\si{\mega\hertz})$ && $261$ & && $205$ &  \\
		$\Delta \omega_{01}$ $(\%)$ && $15$ & && $16$ &  \\
		$E\ped J/E\ped c$ && $179$ & $132$ && $84$ & $61$ \\
		$\omega_{01}$ ($\si{\giga\hertz}$) && $9.66$ & $8.25$ && $5.13$ & $4.30$  \\
		$\left|\chi\right|$ ($\si{\mega\hertz}$) && $1.79$ & $4.09$ && $10.72$ & $2.36$ \\
		$g$ ($\si{\mega\hertz}$) && $82$ & $98$ && $91$ & $61$ \\
		\noalign{\smallskip}\hline
	\end{tabular}
\end{table*} 
\space
\section{The ferro-transmon as a magnetic noise detector}
\label{dissipation}

The high tunability of the ferro-transmon Hamiltonian realizes the ideal playground to study noise fluctuations in ferromagnetic Josephson devices. We show how the combined analysis of relaxation and dephasing processes in the ferro-transmon allows to characterize both magnetization fluctuations and standard  flux noise. Comparing the effects of these two kinds of noise may offer important clues to optimize qubits designs and to understand fundamental aspects of quantum dissipative models.

\subsection{Qubit-noise coupling}

The ferro-transmon Hamiltonian depends on the fluxes $\Phi\ped Z$ and $\Phi\ped L$.  We assume that magnetization fluctuations yield local flux fluctuations, $\delta\Phi\ped L$, while $\Phi\ped Z$-fluctuations,  $\delta \Phi\ped Z$, are dominated by external electromagnetic noise, {\sl i.e.} we neglect orbital effects of magnetization fluctuations.

The fluctuations of $\Phi\ped Z$ and $\Phi\ped L$  in turn lead to fluctuations of two classical parameters: $\vf_0$, defined in Appendix \ref{App1}, and $E_{\rm J}$, defined by Eq.~(\ref{EJ}). To describe their effect on the qubit Hamiltonian ${\cal H}_{\rm Q}$, given by Eq.~\eqref{qubit}, we expand ${\cal H}_{\rm Q}$ to first order in the fluctuation amplitudes, arriving at the following expression for the ferro-transmon - noise coupling Hamiltonian ${\cal H}\ped V(t)$,
\bea \label{Hnoise1} {\cal H}_{\rm V}(t)&=&-\sum_{m={\rm L,Z}}\! \delta\Phi_m \lf[\frac{\partial E_{\rm J}}{\partial \Phi_m} \cos(\vf-\vf_0)+ \rg.\nn\\& & +\lf.\!E_{\rm J}\frac{\partial\vf_0}{\partial \Phi_m}  \sin(\vf-\vf_0)\rg]. \eea
The effect of the different contributions to ${\cal H}\ped V(t)$ can be easily understood switching to the ferro-transmon eigenstate basis. In this basis, keeping only the first two levels and assuming 
$E_{\rm J}(\Phi\ped Z,\Phi_{\rm L})\gg E_{\rm c\Sigma}$, we can write ${\cal H}_{\rm Q}\simeq\omega_{01} \sigma_z$, with $\omega_{01}=\omega_{\rm Q}(\Phi\ped Z,\Phi_{\rm L})-E_{{\rm c \Sigma}}$ and 
\be \label{Hnoise2} {\cal H}_{\rm V}(t)\simeq-\!\!\!\!\!\sum_{m={\rm L,Z}}\!\!\!\lf[ A_{m \parallel} \sigma_z+A_{m\perp}\sigma_y\rg] \frac{\pi\delta\Phi_m}{\Phi_0},\ee
where $A_{m\parallel}$ and $A_{m \perp}$ are respectively given by
\be
A_{m\parallel}=\frac{\Phi_0}{2\pi}\sqrt{\frac{8E_{\rm C\Sigma}}{E_{\rm J}}} \frac{\partial E_{\rm J}}{\partial \Phi_m} \label{Amparallel}
\ee
and 
\be A_{m \perp}=\frac{E_{\rm J}\Phi_0}{\pi}\lf[2\lf(\frac{2E_{\rm c\Sigma}}{E_{\rm J}}\rg)^{1/4}\!\!\!\!\!-\lf(\frac{2E_{\rm c\Sigma}}{E_{\rm J}}\rg)^{3/4}\rg]\frac{\partial\vf_0}{\partial \Phi_m}.\label{Amperp}\ee

Starting from Eqs.~(\ref{Hnoise2})-(\ref{Amperp}) and employing Bloch-Redfield theory~\cite{Vool2017}, we see that, as expected,  fluctuations of $E\ped J$ yield qubit's dephasing, while fluctuations of the phase $\varphi_0$ lead to relaxation.
Therefore, since  both $\delta \Phi\ped Z$ and $\delta \Phi\ped L$ induce both $\varphi_0$ and $E\ped J$ fluctuations, either of them yields both dephasing and relaxation. 

In particular, the fluctuations of the local flux $\delta \Phi_{\rm L}$ can be related  to $H\ped p$ and $M$ fluctuations.
Using Eq.~(\ref{eq:phiL}), $\delta \Phi_{\rm L}$  can be indeed cast as follows
\be\label{eq:dphil}
\delta \Phi\ped L=d_{\rm m} a\,\delta H\ped p+ 4\pi d_{\rm F}a\,\delta M.
\ee
Here, $\delta M$ and $\delta H\ped p$ are not independent. However, by assuming an instantaneous response of the ferromagnet to the pulsed magnetic field fluctuations, we can  write 
\be \label{eq:dm}\delta M(t)=\chi\, \delta H\ped p(t)+\delta M_i(t)\ee  
where $\chi$ denotes the ferromagnet's susceptibility and $\delta M_i$ indicates the fluctuations of the magnetization at $H\ped p=0$ {\sl i.e.} such that $\la \delta M_i\delta H\ped p\ra=0$.
The  assumption of instantaneous response is justified for the purpose of calculating the rates if the ferromagnet magnetization dynamics is faster than the magnetic field pulse sequence time-scale and the qubit's dynamics.

In the following starting from the above equations we calculate the relaxation rates and the dephasing characteristic and we show that it is  possible to isolate the effects of the different noise sources at specific working points.

\subsection{Relaxation}

The total decay rate due to relaxation/excitation processes is given by $\Gamma_1=\Gamma_\downarrow+\Gamma_\uparrow$, where  relaxation (excitation) $\Gamma_\downarrow$  ($\Gamma_\uparrow$) rates in terms of the spectral noise function read as
\be\Gamma_\downarrow=\sum_{m=L,Z} A^2_{m\perp}S_{\Phi_m}(\omega_{01}) \frac{\pi^2}{\Phi_0^2}\ee
\be\Gamma_\uparrow=\sum_{m=L,Z} A^2_{m\perp}S_{\Phi_m}(-\omega_{01}) \frac{\pi^2}{\Phi_0^2}, \ee
where  $S_{X}(\omega)$ denotes the noise spectral function at frequency $\omega$, $S_{X}(\omega)=\large\int \la X(t)X(0)\ra e^{i\omega t}{\rm d}t$.
Specifically,  using Eq.~(\ref{eq:dm}),  $S_{\Phi\ped L}(\omega)$ can be written as follows 
\be \label{eq:s2fi} S_{\Phi\ped L}(\omega)=\beta_1^2 S_{H\ped p}(\omega)+(4\pi\beta_2)^2S_{M_i}(\omega),\ee 
where  $\beta_1=d_m a (1+ \frac{d_F}{d_m}\chi)$ and $\beta_2= d_F a$. To estimate $S_{\Phi\ped Z}$, we follow Refs.~\cite{Ithier2005,Koch2007,Yan2016}, and we consider both the Ohmic and the $1/f$ contributions, respectively denoted as ${S_{\Phi\ped Z}}\api{Ohmic}$ and ${S_{\Phi\ped Z}}^{1/f}$. The former is estimated as~\cite{Ithier2005,Koch2007}
\begin{equation}
\label{noisePhiz}
{S_{\Phi\ped Z}}\api{Ohmic}=L\ped m^2 S_{I_{\rm n}}(\omega_{01}),
\end{equation}
where $L\ped m$ indicates the mutual inductance between the SQUID and the flux-bias loop, and $S_{I_{\rm n}}$ denotes the current noise spectrum in the flux bias circuit. The $1/f$ contribution is modelled as suggested in Ref.~\cite{Yan2016} as: 
\begin{equation}
\label{noisePhiz1f}
{S_{\Phi\ped Z}}^{1/f}=A^2_{\Phi\ped Z}\left(2\pi\frac{\SI{1}{\hertz}}{\omega}\right)^{0.9},
\end{equation}	
with $A_{\Phi\ped Z}=1.4 \mu \phi_0$. As illustrated in Ref.~\cite{Yan2016}, at frequencies of the order of $\omega_{01}$, the $1/f$ contribution dominates over the Ohmic one due to the smallness of the inductance $L\ped m$.
\begin{figure}	
	\begin{center}
		\subfloat[][]{\includegraphics[width=0.9\columnwidth]{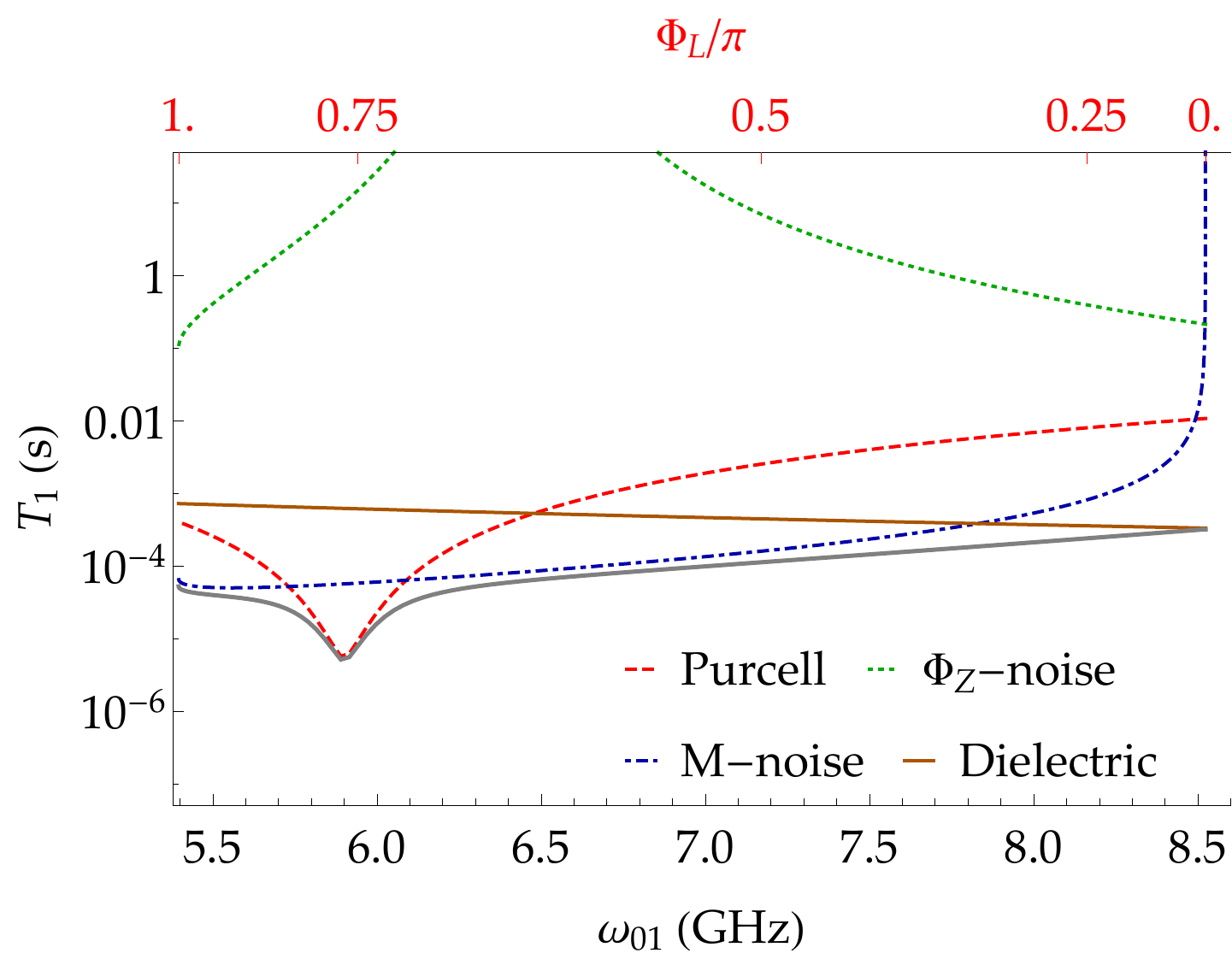}}\\
		\subfloat[][]{\includegraphics[width=0.9\columnwidth]{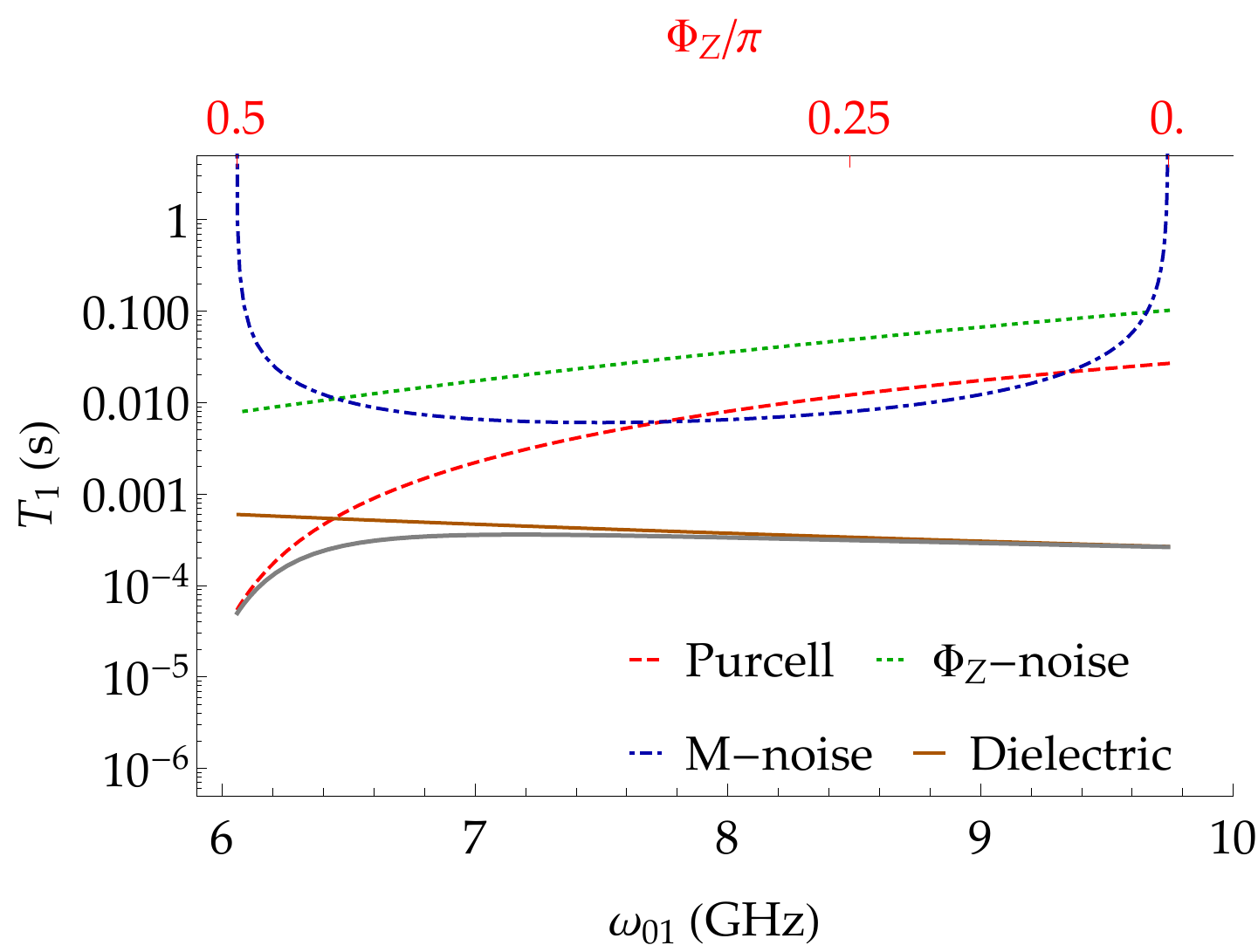}}\\
		\caption{Relaxation time $T_1$ due to Purcell (red dashed curve) and dielectric (orange curve) effects, magnetization (blue dashed-dotted curve) and flux noise (green dotted curve) fluctuations for the ferro-transmon in the hybrid DC-SQUID configuration as a function of the qubit frequency $\omega_{01}$ for $g=\SI{80}{\mega\hertz}$, $E\ped J\api{SFS}(\text{HI})=\SI{32}{\giga\hertz}$, $E\ped J\api{SIS}(\text{HI})=\SI{14}{\giga\hertz}$, $E\ped c=\SI{260}{\mega\hertz}$, $\kappa=\SI{400}{\kilo\hertz}$, $\tau\ped c=\SI{1}{\pico\s}$. In (a), we fix $\phi\ped Z=\pi/4$ and we change $\phi\ped L\in [0,\pi]$, while in (b) we fix $\Phi\ped L\sim\pi/12$ and we change $\phi\ped Z\in [0,\pi/2]$. The grey curve represents the total relaxation time $T_1$, given by the sum of all the relaxation contributions.}
		\label{T1} 
	\end{center}
\end{figure}

In the low-temperature limit $k\ped BT\ll\hbar\omega_{01}$, the above equations eventually allow us to recast the decay rate as the sum of three contributions, {\sl i.e.} $\Gamma_{1}=\Gamma_{1,\Phi\ped Z}+\Gamma_{1,H\ped p}+\Gamma_{1,M_i}$ where  $\Gamma_{1,H\ped p}$ accounts for the fluctuations of the pulsed magnetic field, 
\be  \label{gamma1hp} \Gamma_{1,H\ped p}\simeq A_{L\perp}^2 \beta_1^2 S_{H\ped p}(\omega_{01})/(4\Phi_0^2),\ee
$\Gamma_{1,M_{\rm i}}$ accounts for the fluctuations of the magnetization 
\be\label{gamma1mi}
\Gamma_{1,M_i}\simeq A_{L\perp}^2 (4\pi\beta_2)^2 S_{M_{ i}}(\omega_{01})/(\Phi_0^2)\ee
and $\Gamma_{1,\Phi\ped Z}$ is associated to the fluctuations of the flux $\Phi\ped Z$,
	\begin{equation}
	\label{gamma1hx}
	\Gamma_{1,\Phi\ped z}\simeq {A\ped z}_\perp^2(S_{\Phi\ped Z }\api{Ohmic}+S_{\Phi\ped Z}^{1/f})/\Phi_0^2. 
	\end{equation}
Interestingly, at $H\ped p\equiv 0$, the contributions of  $M\ped i$ and  $\Phi\ped Z$  can be isolated by appropriately choosing the qubit's working point.
Indeed,  according to Eq.~(\ref{Amperp}),  the coupling of $M\ped i$ and  $\Phi\ped Z$ fluctuations are proportional to the derivatives  $\partial \varphi_0/\partial \Phi\ped L$ and $\partial \varphi_0/\partial \Phi\ped Z$, respectively. 
The latter have the following expressions in terms of the Josephson energy  ${E\ped{J}}\api{SFS}$ and the asimmetry parameter $d$,
\be\label{phi0phiL}
\frac{\partial \vf_0}{\partial \phi_{\rm L}}=\frac{1}{4}\frac{(1-d^2)\sin(2\phi_{\rm Z})}{\cos^2(\phi_{\rm Z})+d^2 \sin^2(\phi_{\rm Z})}\frac{\partial \log {E\ped{J}}\api{SFS}}{\partial \phi_L}
\ee
\be\label{phi0phiz}
\frac{\partial \vf_0}{\partial \phi_{\rm Z}}=\frac{d^2}{\cos^2(\phi_{\rm Z})+d^2 \sin^2(\phi_{\rm Z})},
\ee
with $\phi_m=\pi \Phi_m/\Phi_0 $, and $m=\{\text{L,Z}\}$.
We see that  for $d\sim0$, it is possible to suppress $\partial \varphi_0/\partial \phi\ped Z$, and the relaxation rate can be directly related to the intrinsic magnetization noise spectrum $S_{M_{\rm i}}(\omega_{01})$. 
On the contrary, at $\phi_{\rm Z}=0$ and $d\neq 0$, the relaxation rate yields indications on  $S_{\Phi\ped Z}(\omega_{01})$.
Clearly, other relaxation channels such as Purcell effect, dielectric losses and quasi-particle relaxation may contribute to $\Gamma_1$ and hinder the measurement of magnetic fluctuation noise. 

To illustrate this point, in Fig.~\ref{T1} we compare the relaxation induced by magnetization and flux fluctuations, dielectric losses and Purcell effect by plotting the corresponding relaxation times. In order to calculate the relaxation rates associated with magnetization and flux noise in Fig.~\ref{T1}, we use Eqs.~(\ref{gamma1mi}-\ref{gamma1hx}) and we estimate the spectral function $S_{\Phi\ped Z}(\omega_{01})$ using Eq.~(\ref{noisePhiz}), {\sl i.e.} we consider ohmic and $1/f$-type $\Phi\ped Z$-fluctuations~\cite{Koch2007}. Instead, for the spectral function $S_{\Phi\ped L}(\omega_{01})$ we assume that: (i) the amplitude of the magnetization fluctuations satisfy the relation $\delta M_i \lesssim 5\cdot10^{-3} M_{\rm sat}$, with $M_{\rm sat}$ the saturation magnetization of the ferromagnet, and (ii) the magnetization correlation function decays exponentially on a time scale $\tau \ped c \sim \SI{1}{\pico\s}<\tau$, where $\tau$ is the magnetic field pulse sequence time-scale. The corresponding noise spectrum reads as:
\begin{equation}
	S_{\Phi\ped L}(\omega)=\delta M_i^2 \tau\ped c/(\tau \ped c^2 \omega^2+1).
\end{equation}

As reported in Ref.~\cite{Smith2019}, capacitive losses relaxation contribution ${\Gamma_1}\ped c $ are defined as:
\begin{equation}
{\Gamma_1}\ped{c}=\frac{2}{\hbar}\left|\left<0|2eN|1\right>\right|^2\frac{1}{C Q(\omega_{01})},
\end{equation}
where $C$ is the lossy capacitance of the superconducting island in the transmon, with quality factor $Q(\omega)=\text{Im}Y\ped{cap}(\omega)/\text{Re}Y\ped{cap}(\omega)$, defined in terms of the capacitance admittance $Y\ped{cap}(\omega)$~\cite{Smith2019}. Therefore, a crucial role is played by the superconducting and dielectric materials involved in the circuit. In the following, we will refer to standard parameters reported in literature~\cite{Casparis2018,Smith2019,Bilmes2020}.

Finally, to describe Purcell relaxation across different regimes we use the expression given {\sl e.g.} in Ref.~\cite{Sete2014},
\begin{equation}
	{\Gamma_1}\ped{Purcell}=\frac{\kappa}{2}\left(1-\frac{\Delta^2}{\Delta^2+4g^2}\right),
	\end{equation}
where $\kappa=\omega\ped{RO}/Q\ped c$ is the resonator decay rate, $\Delta$ is the qubit-resonator detuning and $g$ is the qubit-resonator coupling. 

In Fig.~\ref{T1} (a) and (b), we show the total relaxation time $T_{1}=1/\Gamma_1$ and the relaxation times corresponding to the magnetization and flux fluctuations, dielectric losses and Purcell noise, respectively given by $1/\Gamma_{1,M_i}$,  $1/\Gamma_{1,\Phi\ped Z}$, $\Gamma_{1,\text{c}}$ and $1/{\Gamma_1}\ped{Purcell}$, as a function of the ferro-transmon qubit frequency $\omega_{01}$. In Fig.~\ref{T1} (a), $\omega_{01}$ is varied by tuning the local flux field $\Phi\ped L$ (red axis in the panel) and setting $\Phi\ped Z=\Phi_0/4$, while in Fig.~\ref{T1} (b) it is varied by tuning $\Phi\ped Z$  and setting $\Phi\ped L=\Phi_0/12$.
	
We see that by properly choosing the flux and local field working point, the ferro-transmon might be used as a magnetic noise detector. For example, in Fig.~\ref{T1} (a) magnetization fluctuations are the dominant relaxation channel in a wide range of local flux field below $\Phi\ped L=0.4\pi$, with the exception of a narrow region occurring for resonant qubit-resonator coupling, where the Purcell effect is the dominant relaxation channel. Above $\Phi\ped L=0.4\pi$, instead, dielectric losses are the dominant relaxation processes.
 
Analogously, dielectric losses are also dominant in Fig.~\ref{T1} (b) in the whole frequency range.  Thus, while magnetization fluctuations can be evaluated by measuring the ferro-transmon relaxation rate, flux-noise fluctuations are expected to be small compared to other relaxation channels~\cite{Koch2007}.  However, it has been demonstrated that a careful choice of the materials into play can minimize this effect, thus allowing to highlight $\Phi\ped Z$-induced relaxation~\cite{Place2021} for specific values of $\Phi\ped Z\sim0$ as predicted by Eq.~\eqref{phi0phiz}.

\subsection{Dephasing}

To describe the dephasing behavior, we start by expressing the density-matrix $\rho_{01}$ in terms of $\delta M_i$, $\delta H\ped p$ and $\delta \Phi\ped Z$ by using Eq.~(\ref{eq:dphil}). Within the assumption of Gaussian noise~\cite{Ithier2005} and free-evolution, it reads as
\bea\label{eq:r012}
\rho_{01}(t)&=& \rho_{01}(0)e^{i\omega_{01}t}\exp\lf[-\frac{A_{Z\parallel}^2}{8\Phi_0^2} \bigg\la\lf( \int_0^t \delta \Phi\ped Z\rg)^2\bigg\ra\rg]\cdot \\
& & \!\!\!\!\cdot  \exp\lf[-\frac{A_{L\parallel}^2}{8\Phi_0^2} \bigg\la \lf(\int_0^t \beta_1 \delta H\ped p+4\pi\beta_2 \delta M_i\rg)^2\bigg\ra\rg]\nn
\eea
where we assumed $\la  \delta \Phi\ped Z \delta \Phi_{\rm L}\ra=0$. The decay characteristic of the qubit dictated by this equation depends sensitively on the noise source spectrum. Specifically, setting
$\la  \delta H\ped p \delta M_i\ra=0$ and  introducing the functions  $f_\lambda(t)=\int_{-\infty}^\infty S_{\lambda}(\omega)\frac{\sin^2(\omega t/2)}{(\omega/2)^2} d\omega$ with $\lambda=\{M_i,\,H\ped p, \Phi\ped Z\}$, we can write
\bea\label{eq:r013}
\rho_{01}(t)&=& \rho_{01}(0)e^{i\omega_{01}t}\exp\lf[-\frac{A_{Z\parallel}^2}{8\Phi_0^2} f_{\Phi\ped Z}(t)\rg]\cdot \\
& & \!\!\!\!\cdot 
\exp \lf[- \frac{A_{L\parallel}^2}{8\Phi_0^2}\lf(\beta_1^2 f_{H\ped p}(t)+(4\pi \beta_2)^2f_{M_i}(t)\rg)\rg].\nn
\eea

In order to calculate the functions  $f_\lambda(t)$ we need information on the typical time scales of the system and of the noise sources. Low-frequency magnetic flux noise in superconducting qubits typically has a $1/\omega^\nu$ with $\nu\in [0.7,1.2]$ and it may be affected by a wide range of factors, including fabrication details, materials properties as well as the structure of the junction and the circuit~\cite{Ithier2005}. Analogously, spontaneous magnetization fluctuations in a ferromagnet due to domain wall dynamics, trapped ions and defect at the ferromagnet surface~\cite{Bruno1990,YAMAURA2001,Vardimon2016} may display different spectral behaviors, such as a $1/\omega$ frequency dependence as reported in Refs.~\cite{gijs1997,ocio1986} or a $1/\omega^\nu$ with $\nu=3/2$ as recently demonstrated by Balk et al. in Ref.~\cite{balk2018}.
\begin{figure}	
	\begin{center}
		\includegraphics[width=0.9\columnwidth]{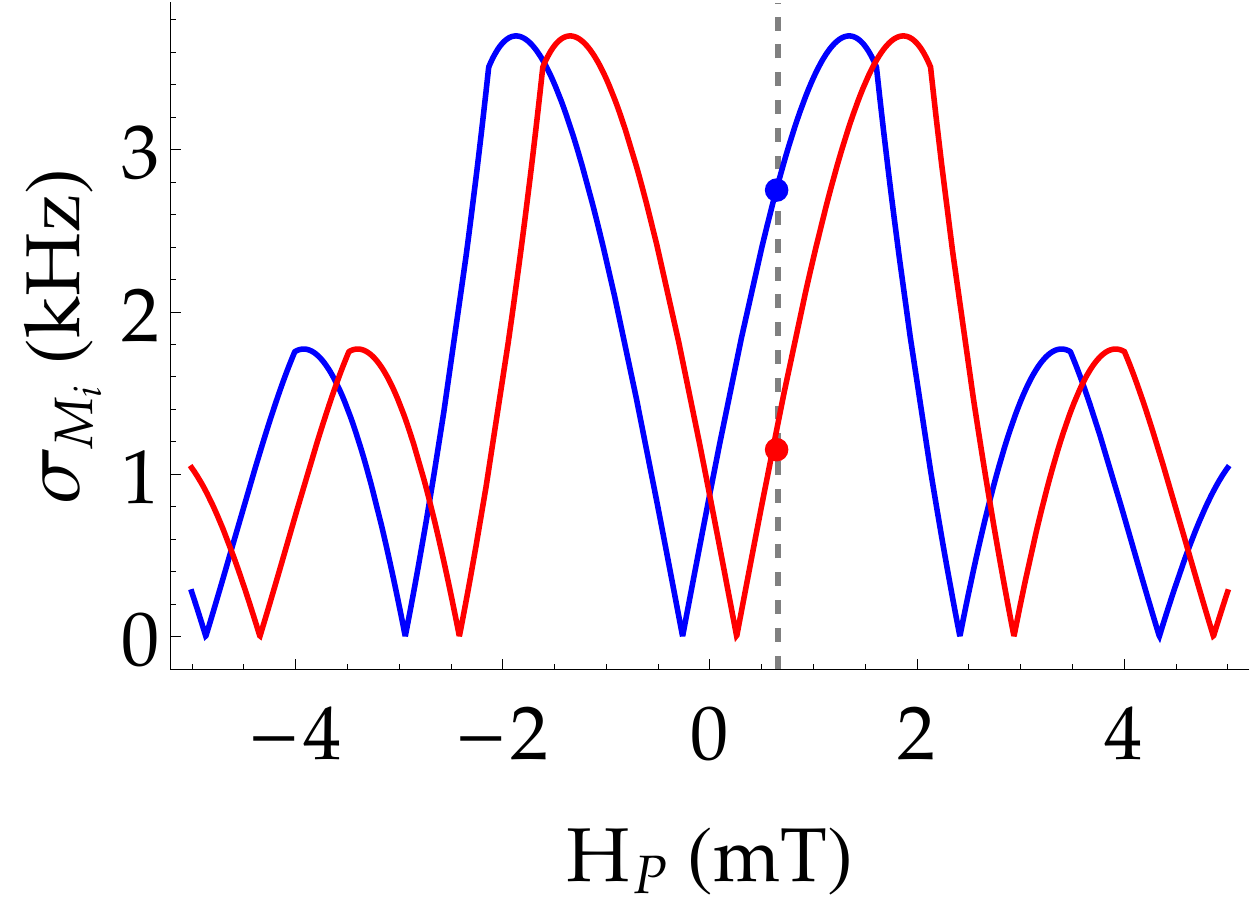}
		\caption{Gaussian decay rate $\sigma_{M_{\rm i}}$ (Eq.~(~\ref{app1})) along the up and down magnetization curves as a function of the local magnetic field.  The red and blue dots correspond to $E\ped J\api{SFS}=\SI{32}{\giga\hertz}$ and \textbf{$E\ped J\api{SFS}=\SI{20}{\giga\hertz}$} and we set $E_{\rm c}=\SI{0.261}{\giga\hertz}$, ${E\ped J}\api{SIS}=\SI{15}{\giga\hertz}$ and $A_{M_{\rm i}}\sim \SI{1}{\micro\tesla}$. The dependence of $E\ped J\api{SFS}$ on the magnetic field is estimated from the Fraunhofer-like curves given in Fig.~\ref{pattern} (b).}
		\label{decay-mag} 
	\end{center}
\end{figure}
We start by deriving a general analytical expression for the decay of $\rho_{01}$  considering only the effect of magnetization fluctuations. We here assume that
\be\label{eq:omeganu}
S_{M_{\rm i}}(\omega)=A_{M_{\rm i}}^2/\omega \cdot (\omega_0/\omega)^{\nu-1} \quad {\rm for}\quad \omega\in[\omega_{\rm ir}, \,\omega_{\rm uv}],
\ee
where $\omega_{\rm ir}$ and $\omega_{\rm uv}$  are the infrared and ultra-violet cut-off, respectively. We also assume that the typical evolution time $t$ satisfies $\omega^{-1}_{\rm uv}<t<\omega^{-1}_{\rm ir}$ and $1/2\leq\nu\leq3/2$  following Refs.~\cite{anton2012,kempf2016}. Starting from the spectrum of Eq.~(\ref{eq:omeganu}), the integral in Eq.~(\ref{eq:r013}) can be easily calculated numerically as shown in App.~\ref{App2}.
\begin{figure*}	
	\begin{center}
		\includegraphics[scale=0.4]{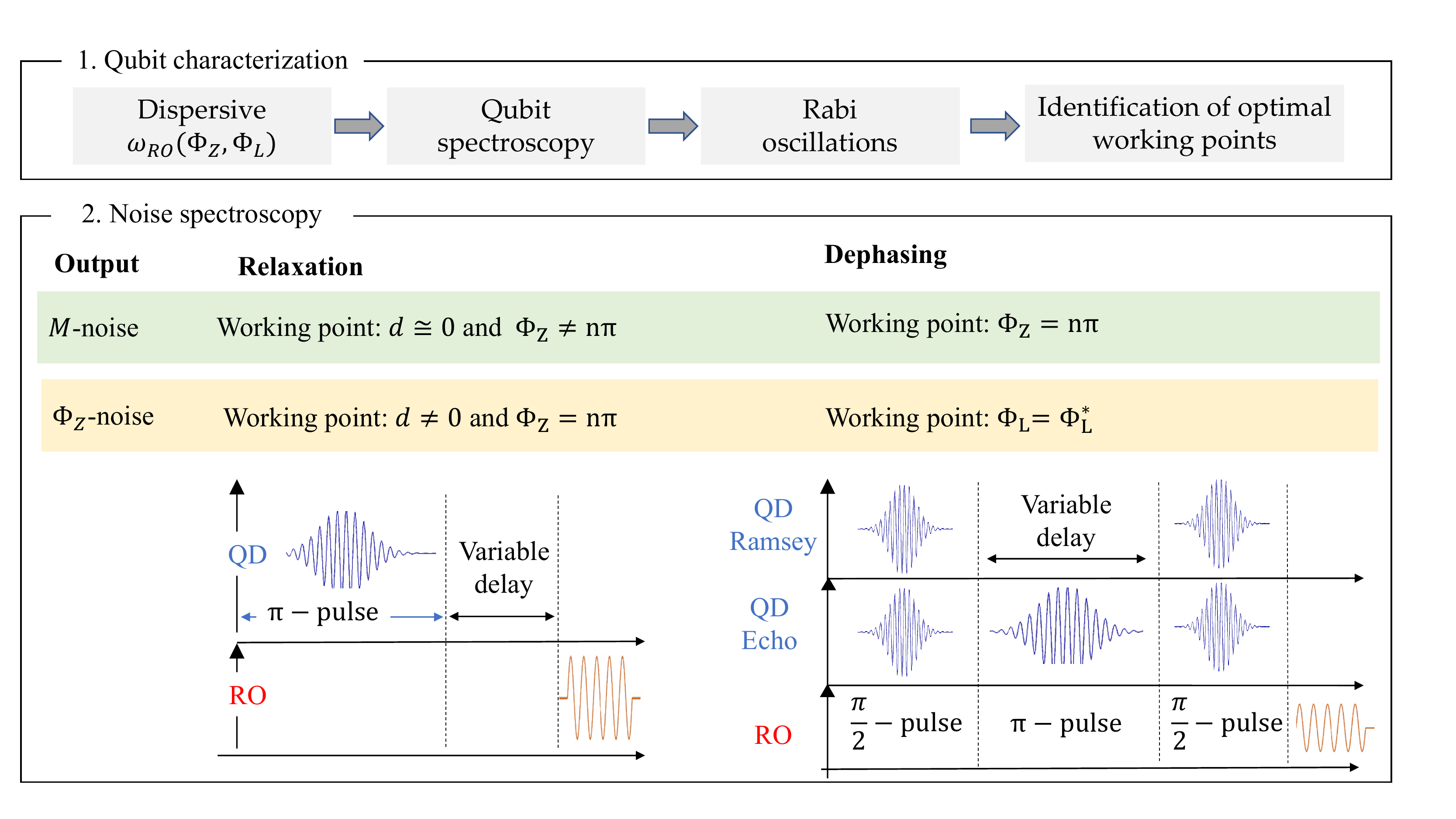}
		\caption{Measurement protocol for the investigation of low- and high-frequency spectral noise functions of the intrinsic magnetic fluctuations (green boxes) and external magnetic field fluctuations (orange boxes). The protocol is divided into two phases: phase (1) regards standard qubit characterization; phase (2) summarizes the study of relaxation phenomena through relaxation time measurements and dephasing time measurements for noise spectroscopy. In phase (2) we also report standard microwave pulsed sequences for the qubit drive (QD) and the read-out (RO) for the relaxation, Ramsey and Echo sequences, respectively~\cite{Krantz2019}.
		}\label{table-noise}
	\end{center}
\end{figure*}
Thus, we can derive simple analytic approximations for $\rho_{01}(t)$. Specifically, for $\nu=1$ we obtain
\be\label{app1}
\rho_{01}(t)\propto e^{-{(\sigma^2_{M_{\rm i}} t^2)\ln(\pi/(\omega_{\rm ir }t))}} \quad {\rm with}\quad \sigma_{M_{\rm i}}\simeq  A_{L\parallel}\frac{ \pi \beta_2 A_{M_{\rm i}}}{\Phi_0},
\ee
while for $\nu \neq 1$ we can write $\rho_{01}(t)\propto \exp^{-\sigma^2_{M_{\rm i}} t^2 \alpha_{\nu}(t)} $ with
\be\label{app2}
\alpha_\nu(t)=\frac{1-\lf(\omega_{\rm ir}t/\pi\rg)^{\nu-1}}{\nu-1} \lf(\frac{\omega_0}{\omega_{\rm ir}}\rg)^{\nu-1},
\ee
valid for $\omega_{\rm ir}t \ll 1$.

For the other sources of $1/f$ noise we can make similar reasoning and obtain analogous results, {\sl i.e.} taking into account the different coupling constants we can write:
\be
\sigma_{H_{\rm L}}\simeq  A_{L\parallel}\frac{ \pi \beta_1 A_{H_{\rm L}}}{\Phi_0} \quad {\rm and} \quad \sigma_{\Phi\ped Z}\simeq  A_{{\rm Z}\parallel}\frac{ \pi A_{\Phi\ped Z}}{\Phi_0},
\ee
assuming for both noise sources $\nu=1$. For $\Phi\ped Z$-noise a more precise description of the behavior of the spectral function can be given by Eq.~\ref{noisePhiz1f}~\cite{Yan2016}.

For $\omega\in[\omega_{\rm ir}, \,\omega_{\rm uv}]$, where $\omega_{\rm ir}$ and $\omega_{\rm uv}$ may be different for the different noise sources, we made the following assumptions on the external flux and local magnetic field fluctuations:
\be\label{eq:omeganuL}
S_{H_{\rm L}}(\omega)=A_{H_{\rm L}}^2/\omega \cdot (\omega_0/\omega)^{\nu-1} 
\ee
and 
\be\label{eq:omeganux}
S_{\Phi\ped Z}(\omega)=A_{\Phi\ped Z}^2/\omega \cdot (\omega_0/\omega)^{\nu-1}.
\ee

Such low-frequency behavior of magnetic spectral functions is encoded in the dephasing characteristic and can be accessed by Ramsey and  echo experiments~\cite{Koch2007}. Also in this case, the contributions of magnetization and $\Phi\ped Z$ fluctuations can be separated. Indeed at the external flux sweet spots located at $\phi_{\rm Z}=n\pi$ as in standard transmons, the dephasing is mostly due to intrinsic magnetization fluctuations, while the effect of  external magnetic field fluctuations is negligible. On the other hand, for $\phi_{\rm Z}\neq n\pi$  and $\phi_L$ tuned at the local flux sweet spots, $\Phi\ped L=\Phi\ped L^*$, ferro-transmon's dephasing probes  $\Phi\ped Z$'s fluctuations. Notice that the local flux sweet spots, defined by the condition $\frac{\partial E\ped J\api{SFS}}{\partial \Phi\ped L}\big|_{\Phi\ped L=\Phi\ped L^*}\!\!\!=0$, depend on the properties of the SFS JJ.

Also in the case of dephasing rates, we can roughly estimate their order of magnitude. In Fig.~\ref{decay-mag}, we plot the behavior of the gaussian decay rate for the intrinsic magnetic fluctuations ${\sigma\ped{M}}\ped i$, as given by Eq.~(\ref{app1}), and as a function of the local magnetic field. ${\sigma\ped{M}}\ped i$ has been obtained by fitting with a Fraunhofer-like pattern the up and down $I\ped c(H\ped p)$ curves in Fig.~\ref{pattern} (b), and rescaled in order to achieve the suitable $E\ped J\api{SFS}$ in Fig.~\ref{final} (a). The red and blue dots in Fig.~\ref{decay-mag} indicate two possible HI and LO working points corresponding to the red and blue resonances spectra shown in Fig.~\ref{final} (c), respectively. As discussed in Sec.~\ref{highlights}, in principle, we can adapt the pulsed magnetic field sequence in order to reach the largest decay time due to magnetization fluctuation dephasing. 

\subsection{Noise detection protocol}

The analysis of the effect of different noise sources on the ferro-transmon design allows to propose a novel protocol for the study of the spectral noise functions in hybrid ferromagnetic quantum systems. In Fig.~\ref{table-noise} we highlight how we can access to $\Phi\ped Z$ and $M_{\rm i}$ noise spectra. In the first phase, a basic characterization of the qubit is in order to obtain information on the sweet-spots in presence of $\Phi\ped Z$, the maximum of the $E\ped J$ in presence of $\Phi\ped L$ and the points at which the asimmetry between the SIS and the SFS JJs in the loop is zero. This is a fundamental step also for extracting the $\pi$-pulse for the study of relaxation and the dephasing processes in the system. In the second phase, we propose to perform $T_1$ measurements in two configurations: (i)  at $d=0$ and $\Phi\ped Z\neq0$ to address high-frequency magnetization-noise spectra, and (ii) for asymmetry $d\neq0$ and $\Phi\ped Z=0$ to address high-frequency $\Phi_{\rm Z}$-noise spectra. The third step involves standard Ramsey and Echo sequence in other two configurations: (i) at the sweet-spots for $\Phi\ped Z$-noise spectra to address low-frequency $M_{\rm i}$-noise spectra and (ii) in the maximum of the Fraunhofer-like modulation of the Josephson energy to address low-frequency $\Phi\ped Z$-noise. We remark that the simultaneous measurement of relaxation and dephasing is also extremely relevant to fully characterize 1/f noise, as demonstrated  in various works~\cite{Yan2016,Quintana2017,kempf2016}.

\section{Conclusions}

The natural digital behavior of SFS JJs and the recent advancements in fabrication of high-quality and low-dissipative tunnel SFS JJs give the possibility to implement hybrid quantum devices compatible with energy-efficient cryogenic digital electronics. We have discussed the capabilities and the feasibility of a ferro-transmon that uses a hybrid ferromagnetic SQUID coupled to a superconducting resonator, thus giving a first estimation of the electrical and magnetic parameters needed to fabricate a reliable and measurable device. By considering capacitive elements in the circuit design in line with typical values already implemented in non-magnetic transmon devices, we demonstrated the feasibility of the proposed circuit. The insertion of a ferromagnetic barrier in a JJ allows for the investigation of new problems due to the interplay between the two competiting superconducting and ferromagnetic order parameters~\cite{Buzdin2005,Eschrig2008}. Beside its potential for quantum computing applications, a hybrid ferromagnetic qubit in turn offers the possibility to study the dynamics of the tunnel-SFS JJ itself, yielding novel experimental tools to probe quantum phenomena at the interface. We demonstrate that investigating qubit's dephasing and relaxation at suitable working points it is possible to probe both magnetization fluctuations and magnetic flux noise.

\begin{acknowledgements}

The work has been supported by the project “EffQul - Efficient integration of hybrid quantum devices” - Ricerca di Ateneo Linea A, CUP: E59C20001010005 and the project “SQUAD - On-chip control and advanced read-out for superconducting qubit arrays”, Programma STAR PLUS 2020, Finanziamento della Ricerca di Ateneo, University of Napoli Federico II. This work was also co-funded by European Union - PON Ricerca e Innovazione 2014-2020 FESR/FSC - Project ARS01\_00734 QUANCOM.  H.G.A., D.M., D. Mo. and F.T. thank NANOCOHYBRI project (COST Action CA 16218). The authors also thank Mark G. Blamire and Avradeep Pal for useful discussions.

\end{acknowledgements}

\section*{Conflict of interest}

The authors declare that they have no conflict of interest.

\appendix

\section{Ferro-transmon Hamiltonian}
\label{App1}

We here highlight the derivation of the ferro-transmon charging Hamiltonian.
Following the standard approach discussed {\sl e.g.} in Ref.~\cite{Vool2017} we can easily derive the charging Hamiltonian of the ferro-transmon starting from its Lagrangian:
\begin{equation}
{\cal L}\ped c(\dot{\Phi\ped r}, \dot{\Phi})=\frac{C\ped r\dot{\Phi\ped r}^2}{2}+\frac{C\ped b\dot{\Phi}^2}{2}+\frac{C\ped g}{2}(\dot{\Phi\ped r}-\dot{\Phi})^2.
\end{equation} 
where $\Phi\ped r$  and $\Phi$ denote respectively the resonator and qubit node fluxes. 

Starting from the above equation, we can obtain an expression for the charging Hamiltonian written in term of the charges $Q_{\rm r}$ and $Q$, conjugate to the fluxes $\Phi_{\rm r}$ and $\Phi$, and  defined as usual as  $Q_i=\partial_{\dot{\Phi_i}}{\cal L}\ped c(\dot{\Phi_i})$.
By doing so, we obtain
 $ {\bf Q}=\hat M^{-1}\mathbf{\dot\Phi}$, with $\hat M^{-1}$ denoting  the capacitance matrix 
 \begin{equation}
\hat M^{-1}=\left(\begin{matrix}
C\ped r+C_g & -C\ped g\\
-C\ped g & C\ped b+C_g\\
\end{matrix}\right),
\end{equation}  
 and 
\begin{equation}
\label{hamiltonian}
\mathcal{H}_{\rm c}=\frac{1}{2} {\bf Q}\hat M{\bf Q}.
\end{equation}
By inverting the capacitance matrix, we can thus directly obtain information on the interplay between the capacitive elements in the circuit and their effect on the total energy of the system~\cite{Koch2007}. Let us now discuss more in details the different contributions to the ferro-transmon Hamiltonian.

Given the small dimensions of standard transmon devices ($\sim \SI{100}{\micro\m^2}$)~\cite{Walraff2004,Majer2007}, the transmission line resonator in the circuit can be described by a harmonic LC oscillator in the lumped element limit. Hence, the resonator Hamiltonian $H\ped r$ reads as
\begin{equation}
\label{resonator}
\mathcal{H}\ped r=\frac{Q\ped r^2}{2{C_{\Sigma}}\ped r}+\frac{\Phi\ped r^2}{2L\ped r},
\end{equation}
where ${C_{\Sigma}}\ped r$ is the effective total capacitance of the resonator circuit net, which depends on all the capacitance of the circuits, and reduces to $C\ped r$ in the limit $C\ped r\gg C\ped b,C\ped g$~\cite{Koch2007}. The coupling between the qubit and the resonator, $\mathcal{H}\ped{coupl}$, can be cast as 
\begin{equation}
\mathcal{H}\ped{Q-r}=\frac{Q Q\ped r}{2{C_{\Sigma}}\ped{Coupl}},
\end{equation}
where $Q$ is the excess charge in the SQUID circuit and ${C_{\Sigma}}\ped{Coupl}$ is the total effective coupling capacitance. These two terms are the same as occurs in standard transmon devices, and derivation of their explicit expressions can be found in Ref.~\cite{Koch2007}. 

The qubit Hamiltonian $\mathcal{H}\ped Q$ can be written as 
\begin{equation}
\label{qubit0}
\mathcal{H}\ped Q=\frac{Q^2}{2C_{\Sigma}}-E \ped{J}\api{SIS} \cos \vf_1-E \ped{J}\api{SFS}(\Phi\ped L) \cos \vf_2
\end{equation}
where  $\vf_1$ and $\vf_2$ are the phase differences across the SIS and SFS junctions respectively, $E \ped{J}\api{SIS} $ and $E \ped{J}\api{SFS}$ are the corresponding Josephson couplings and $C_{\Sigma}$ is the total effective capacitance obtained from Eq.~(~\ref{hamiltonian}),
\begin{equation}
\label{total_capacitance}
C_{\Sigma}=\frac{(C\ped g+C\ped b)C\ped r+C\ped gC\ped b}{C\ped r+C\ped g}.
\end{equation} 
Assuming negligible inductance of the SQUID loop we can set 
$\vf_1-\vf_2=2\pi\Phi\ped Z/\Phi_0$, $\Phi_0$ denoting the elementary flux, and we can  recast the Hamiltonian $\mathcal{H}\ped Q$ as 
\begin{equation}
\label{qubit}
\mathcal{H}\ped Q= \frac{Q^2}{2C_{\Sigma}}-E\ped J(\Phi\ped Z,\Phi_{\rm L})\cos (\vf-\vf_0),
\end{equation}
where $\vf=\vf_1+\vf_2$, and $E\ped J(\Phi\ped Z,\Phi_{\rm L})$ is defined in Eq.~(~\ref{EJ}) in terms of the local pulsed flux $\Phi_{\rm L}$ and the external flux $\Phi\ped Z$. The phase-shift $\varphi_0$ also depends on $\Phi_{\rm L}$ and the external flux $\Phi\ped Z$ through the asimmetry parameter in Eq.~(~\ref{d}), and reads as
\begin{equation}
\label{phi0}
\tan \vf_0(\Phi\ped Z,\Phi_{\rm L})= -d(\Phi_{\rm L})\tan \Phi\ped Z.
\end{equation}
In the transmonic regime, $E\ped J(\Phi\ped{Z},\Phi\ped L)\gg E_{c\Sigma}$, ${\cal H}\ped Q$ reduces to the following  Duffing Harmonic oscillator Hamiltonian~\cite{Duffing1918} 
\begin{equation}
\label{quantum}
\mathcal{H}\ped Q \simeq\hbar\omega\ped Q(\Phi\ped{Z},\Phi\ped L)\left(b^{\dagger}b+\frac{1}{2}\right)-\frac{{E\ped c}_{\Sigma}}{12}(b+b^{\dagger})^4,
\end{equation}
where $b\;(b^{\dagger})$ are the annihilation (creation) operators, whose definition follows the standard notation introduced in Ref.~\cite{Koch2007}, $\omega\ped Q(\Phi\ped{Z},\Phi\ped L)$ is the qubit frequency in the harmonic approximation, defined as
\begin{equation}
\omega\ped Q(\Phi\ped{Z},\Phi\ped L)=\sqrt{8E\ped J(\Phi\ped{Z},\Phi\ped L)E\ped c},
\end{equation}
and the quartic term in Eq.~(~\ref{quantum}), yielding  the anharmonicity, depends only on  the circuit's  charging energy~\cite{Koch2007},
\begin{equation}
\label{Ec}
{E\ped c}_{\Sigma}=\frac{e^2}{2C_{\Sigma}}.
\end{equation}
For tunnel SFS JJs, the capacitance is the same order of magnitude of that in non-magnetic tunnel JJs~\cite{Wild2010,Massarotti2015,Ahmad2020}. As a consequence, the charging energy of the circuit is mostly unaffected by the presence of the SFS JJ, and the main difference between the ferro-transmon and a standard transmon circuit is  related to the Josephson energy, as discussed in Sec.~\ref{highlights}. 

\section{Analytic approximations for the ferro-transmon-noise couplings and the decay characteristics}
\label{App2}

\begin{figure}[t!]	
	\begin{center}
		\includegraphics[width=0.35\textwidth]{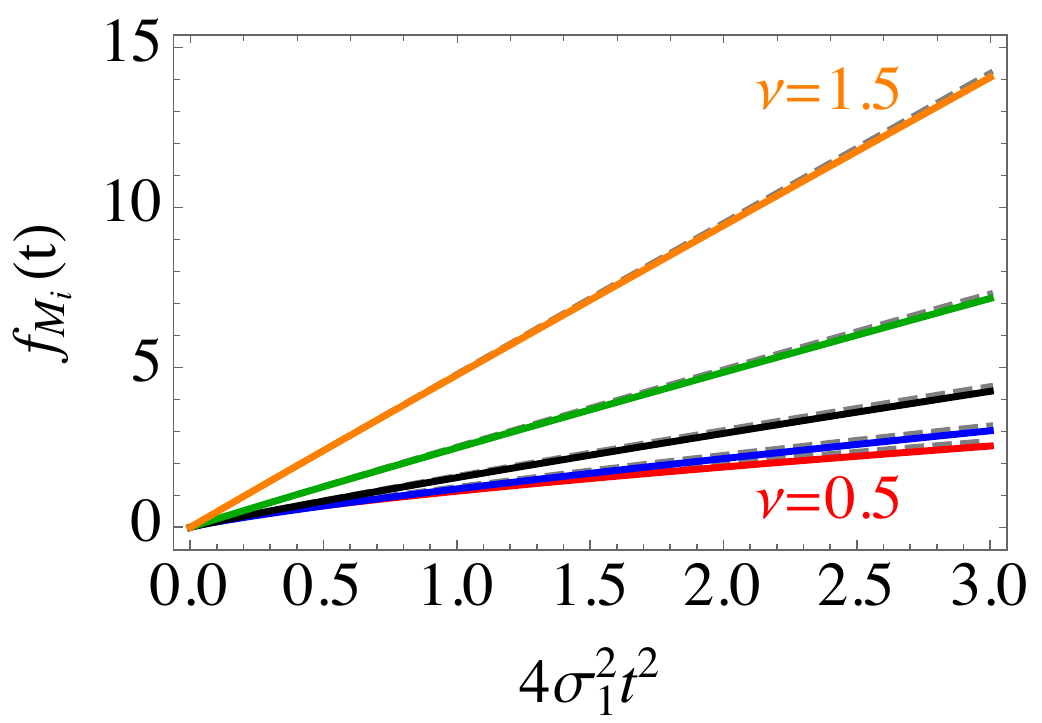}
		\caption{Behavior of the decaying exponent of $\rho_{01}(t)$ for different $1/f$ noise's exponents $\nu=1.5,\,1.25,\, 1,\, 0.75,\,0.5$ and infrared $\omega_{\rm ir}=0.01\,\omega_0$. The gray dashed lines represent the corresponding analytic approximations defined by Eqs.~(~\ref{app1})-(~\ref{app2}).}\label{decay}
	\end{center}
\end{figure}
We outline here the derivation of Eqs.~(~\ref{Amparallel})-(~\ref{Amperp}).
Starting from the Hamiltonian~\eqref{Hnoise1} one easily realizes that in the truncated basis the ferro-transmon noise couplings will be determined by the following matrices:
\bea
\hat A_{mn}&=&\la\psi_m|\cos(\vf-\vf_0)|\psi_n\ra \nn\\
\hat B_{mn}&=&\la\psi_m|\sin(\vf-\vf_0)|\psi_n\ra
\eea
with $|\psi_n\ra$ with $n=0,1$ representing the lowest transmon eigenfunctions. The matrices $\hat A$ and $\hat B$  can be calculated employing the harmonic approximation  as follows
$$\hat A_{mn}\simeq \hat A^H_{mn}=\la\psi^H_m|\frac{(\vf-\vf_0)^2}{2}|\psi^H_n\ra$$ and 
$$\hat B_{mn}\simeq \hat B^H_{mn}=\la\psi^H_m|\lf[(\vf-\vf_0)+\frac{(\vf-\vf_0)^3}{6}\rg]|\psi^H_n\ra.$$ 
Using  these approximated expressions  in Eq.~\eqref{Hnoise1}  and using Eq.~\eqref{Hnoise2} we can then write
$$A_{m\perp}=\Tr\lf[ \hat B^H \sigma_x\rg]  \quad {\rm and} \quad
 A_{m\parallel}=\Tr\lf[ \hat A^H \sigma_z\rg],$$
and we can recover Eqs.(~\ref{Amparallel})-(~\ref{Amperp}) by a simple calculation.

To conclude this section we report in Fig.~\ref{decay} the decaying exponent  $f_{M_i}$ of $\rho_{01}(t)$,
\begin{equation}
\rho_{01}(t)=\rho_{01}(0)e^{i\omega_{01}t}e^{-f_{M_i}(t)},
\end{equation}
as a function of $t^2$ for different values of the $1/f$ noise sources exponent $\nu$.
The gray dashed lines in Fig.~\ref{decay} represent the analytic approximations that are in good agreement with the numerical results. As explained in Sec.~\ref{dissipation}, the determination of the exponent $\nu$ for different noise sources can be done by combined relaxation and dephasing measurements at suitable  working points.

%

\end{document}